\documentclass[12pt,a4paper]{article}
\pdfoutput=1
\usepackage{graphicx,epsfig}
\usepackage{dcolumn} 
\usepackage{slashed,color,amsmath,amssymb}
\usepackage{jheppub}
\usepackage{enumitem}
\usepackage{siunitx}
\usepackage{bm}
\usepackage{layouts}
\usepackage[svgnames]{xcolor}
\usepackage{float}
\usepackage{caption, subcaption}
\usepackage{multirow}
\usepackage{verbatim}
\usepackage[normalem]{ulem}
\newcommand{\be}{\begin{equation}}
\newcommand{\ee}{\end{equation}}
\newcommand{\bea}{\begin{eqnarray}}
\newcommand{\eea}{\end{eqnarray}}
\newcommand{\bit}{\begin{itemize}}
\newcommand{\eit}{\end{itemize}}

\newcommand{\lam}{\lambda}
\newcommand{\kap}{\kappa}

\definecolor{greeen}{HTML}{008ae6}

\definecolor{newgreen}{HTML}{009900}
\definecolor{newnewgreen}{HTML}{21db2c}

\newcommand\thefontsize{The current font size is: \f@size pt}

  \definecolor{fsred}{HTML}{c71616}

\definecolor{fsblue}{HTML}{1e88e5}

\definecolor{fsyellow}{HTML}{dca607}

\definecolor{fsgreen}{HTML}{014d40}

\definecolor{fsviolet}{HTML}{a11b9f}

\definecolor{newpurple}{HTML}{7B1DAA}

\def\gsim{\lower0.5ex\hbox{$\:\buildrel >\over\sim\:$}}
\def\lsim{\lower0.5ex\hbox{$\:\buildrel <\over\sim\:$}}

\hypersetup{
   colorlinks=true,       
   linkcolor=blue,        
   citecolor=red,         
   filecolor=magenta,     
}

\definecolor{newpurple}{HTML}{7B1DAA}

\preprint{\begin{flushright} BONN-TH-2026-16
	\end{flushright}}	

\graphicspath{{./figs/}}

\title{The ABC of RPV III: Classification of R-parity Violating 
Signatures from LQD Couplings and their Coverage at the LHC}

    \author[a]{Herbi\,K.\,Dreiner,}
	\emailAdd{dreiner@uni-bonn.de}
	\affiliation[a]{Bethe Center for Theoretical Physics \& Physikalisches Institut der Universit\"at Bonn,\\ Nu{\ss}allee 12, 53115 Bonn, Germany}

	\author[b]{Yong Sheng Koay,}		 
    \emailAdd{yongsheng.koay@physics.uu.se}
    \affiliation[b]{Department of Physics and Astronomy, Uppsala University, Sweden}		
    
    \author[c]{Tina Potter,}
    \emailAdd{tinapotter@gmail.com}
    \affiliation[c]{Cavendish Laboratory, University of Cambridge, Cambridge, United Kingdom}	
		
    \author[a]{Martin Sch\"urmann,}
	\emailAdd{marschu@uni-bonn.de}
					
	\author[a]{Rhitaja Sengupta,}
	\emailAdd{rsengupt@uni-bonn.de}
		
	\author[a]{Apoorva Shah,}
	\emailAdd{ashah@uni-bonn.de}

    \author[d]{Nadja Strobbe}
	\emailAdd{nstrobbe@umn.edu}
    \affiliation[d]{School of Physics \& Astronomy, University of Minnesota, Minneapolis, MN 55455, USA}
    
\abstract{We consider the R-parity violating Minimal Supersymmetric 
Standard Model (RPV-MSSM) with its wide range of signatures as a test model
for the coverage of beyond the Standard Model searches at the LHC. Here we
present a systematic study of the phenomenological and experimental status 
of specifically the R-parity violating $LQ\bar D$ operators, as
the third study in the ABC of RPV series, following our previous analyses of the $LL\bar E$ and $\bar U\bar D\bar D$ operators. We classify the distinct collider signatures for all possible LSPs and for eight representative $LQ\bar D$ couplings, considering both direct LSP production and production through gauge cascades. Assuming one non-zero $LQ\bar D$ coupling at a time, we assess the current LHC coverage using ATLAS and CMS searches implemented in \texttt{CheckMATE\,2}, and identify recent searches relevant for future recasting. We find substantial sensitivity to the colored sector, while gaps remain for wino- and higgsino-like LSPs for couplings involving $\tau$ leptons, when the LSP is the only sparticle within the LHC kinematic limit. Slepton LSPs remain unconstrained in their direct production at the LHC, although some sensitivity is found for slepton NLSPs with a bino-like LSP. These results highlight gaps in the \texttt{CheckMATE\,2} search database, as well as the need for targeted LHC searches for specific $LQ\bar D$ signatures.
}

\begin{document}
\maketitle




\section{Introduction}
\label{sec:intro}
The Standard Model of particle physics (SM) has been experimentally highly 
successful \cite{ParticleDataGroup:2024cfk}. However, there remain several 
well-known open questions, \textit{e.g.} What is the nature of dark matter 
\cite{Arbey:2021gdg}? What is the origin of the matter asymmetry in the 
Universe \cite{ParticleDataGroup:2024cfk}? And what is the solution to the 
gauge hierarchy problem \cite{Gildener:1976ai, Veltman:1980mj}? The 
answers to these questions all require beyond the SM (BSM) physics. In this
paper we shall consider the RPV-MSSM with its wide range of signatures as a
test model for the coverage of BSM searches at the LHC.

Supersymmetry (SUSY) is a proposed solution to the gauge hierarchy problem 
and addresses the other questions as well. See for example Refs.~\cite{Dreiner:2008tw, Feng:2013pwa, Baer:2020kwz, 
Dreiner:2023yus} for reviews. The most widely discussed supersymmetric 
model is the ``Minimal Supersymmetric Standard Model" (MSSM) 
\cite{Kane:1993td, Drees:2004jm, Baer:2006rs, Dreiner:2023yus}. 
The MSSM has the minimal field content required for a supersymmetric extension of the SM, comprising an additional Higgs doublet and a supersymmetric partner for each SM field, collectively referred to as sparticles, with spins differing by $1/2$ from their SM counterparts.
Furthermore, it restricts the set of
renormalizable couplings by imposing the multiplicative discrete 
symmetry R-parity, 
\begin{equation}
R_p=(\mathbf{-1})^{2S+3B+L}\,,
\end{equation} 
which is also equivalent to the $\mathbb{Z}_2$-symmetry matter parity
\cite{Ibanez:1991hv, Ibanez:1991pr, Dreiner:2005rd, Dreiner:2012ae}. 
The superpotential is given by
\begin{equation}
    W_{\mathrm{MSSM}}= \epsilon_{ab}\left[(\mathbf{Y}_E)_{ij}
    L_i^a H_1^b\bar E_j + (\mathbf{Y}_D)_{ij}
    Q_i^a H_1^b\bar D_j + (\mathbf{Y}_U)_{ij}
    Q_i^a H_2^b\bar U_j -\mu H_1^a H_2^b
    \right]\,.
\label{eq:RPC-Superpot}    
\end{equation}
Here we follow the notation of Ref.~\cite{Allanach:2003eb}. $L_i,\,\bar 
E_i$ denote the leptonic, $Q_i,\,\bar D_i,\,\bar U_i$ the quark, and 
$H_{1,2}$ the Higgs doublet chiral superfields. $i,j\in\{1,2,3\}$ are 
generation indices and $a,b$ are SU(2)$_L$. $\mathbf{Y}_{E,D,U}$ are 
dimensionless Yukawa coupling matrices, $\mu$ is a dimensionful Higgs 
mixing parameter.
$R_p$ prohibits all renormalizable baryon- and lepton-number violating
couplings in the superpotential. This guarantees that the proton is 
stable. R-parity conservation furthermore implies that supersymmetric 
particles can only be produced in pairs, for example at the LHC. It also
implies the lightest supersymmetric particle (LSP) is stable. This 
is typically the lightest neutralino in the MSSM. 

SUSY has been extensively searched for in the context of the 
MSSM at the LHC, to-date to no avail. Typical lower mass bounds on 
strongly interacting supersymmetric particles (squarks, gluinos) are of
the order of a few TeV \cite{ATLAS:2023afl, ATLAS:2025wln}, see also
Refs.~\cite{Canepa:2019hph, ParticleDataGroup:2024cfk} for overviews.
Weaker bounds are obtained for electroweak particles with or without 
compressed mass spectra  \cite{Dreiner:2012sh, CMS:2025ttk, 
ATLAS:2025lhc}. The lightest neutralino can still be massless 
\cite{Dreiner:2009ic, ParticleDataGroup:2024cfk}.

It has been argued that the MSSM with \textit{conserved} R-parity is 
not the only well-motivated minimal supersymmetric model, \textit{i.e.} 
with  the minimal field content \cite{Ibanez:1991hv, Ibanez:1991pr, 
Hempfling:1995wj, Dreiner:2003hw, Dreiner:2003yr, Dreiner:2006xw, 
Dreiner:2011ft, Dreiner:2012ae}. In order to protect the proton, it is 
sufficient to just separately require either baryon- or lepton-number 
conservation \cite{Smirnov:1996bg, Chamoun:2020aft}. This can be 
achieved for example by the $\mathbb{Z}_3$ discrete symmetry baryon 
triality, which is discrete gauge anomaly-free with the minimal field 
content \cite{Ibanez:1991pr, Dreiner:2006xw, Lee:2010vj, Dreiner:2011ft,
Dreiner:2012ae}. Baryon-triality allows for renormalizable lepton-number
violating operators in the superpotential, which we shall consider in 
this paper.

The alternative to the R-parity conserving MSSM (RPC-MSSM) which we 
consider here is the R-parity violating MSSM (RPV-MSSM) \cite{Allanach:2003eb} with the superpotential 
\begin{eqnarray}
W_{\mathrm{TOT}}&=&W_{\mathrm{MSSM}} + W_{\mathrm{RPV}}\,, \\
W_{\mathrm{RPV}}&=& \epsilon_{ab}\left[\lam_{ijk}
    L_i^a L_j^b\bar E_k + \lam'_{ijk}
    L_i^a Q_j^b \bar D_k + \lam''_{ijk} \bar U_i \bar D_j 
    \bar D_k -\kappa_i L_i^a H_2^b
    \right]\,.
\label{eq:RPV-superpot}    
\end{eqnarray}
Here the notation is as in Eq.~\eqref{eq:RPC-Superpot}. The $\lam,\, 
\lam', \,\lam''$ are dimensionless Yukawa couplings and the $\kap_i$
are dimensionful parameters with the leptonic and the Higgs chiral 
superfields. We have dropped SU(3)$_c$ color indices. As mentioned above,
we consider here only lepton-number violation and thus set $\lam''_{ijk} 
=0$. Furthermore as argued in Ref.~\cite{Allanach:2003eb}, at the 
unification scale the $\kap_i$ can be set to zero. At the electroweak 
scale, $\kap_i\not=0$ are generated through the renormalization group
equations \cite{deCarlos:1996ecd, Allanach:1999mh}, however they are 
typically very small and we neglect them here. See also 
Refs.~\cite{Hempfling:1995wj, deCampos:2007bn}. As the parameters 
$\lam,\,\lam',\,\lam''$ violate lepton- or baryon-number, there is a large
collection of bounds on them, typically as a function of a supersymmetric
particle mass \cite{Barger:1989rk, Dreiner:1997uz, Bhattacharyya:1997vv, 
Allanach:1999ic, Chemtob:2004xr, Barbier:2004ez, Dreiner:2012mx, 
Domingo:2018qfg}. 
 
The collider phenomenology of the RPV-MSSM is typically 
very different from the RPC-MSSM \cite{Dreiner:1997uz, 
Barbier:2004ez}. Supersymmetric particles can be singly produced,
and the LSP can decay. The LSP is thus not a dark matter candidate, and
any sparticle can be the LSP \cite{Dreiner:2008ca, Dercks:2017lfq}. 
Furthermore the lightest neutralino can take on any mass 
\cite{Choudhury:1999tn, Gogoladze:2002xp, Dreiner:2009ic}. The collider
phenomenology of the RPV-MSSM has been investigated in some detail from an early stage 
\cite{Hall:1983id, Dawson:1985vr, Dimopoulos:1988fr}. 
In particular, the diverse signatures arising from LSP decays in the RPV-MSSM have been explored in a wide range of phenomenological studies; see, e.g., Refs.\,\cite{Dreiner:1994tj,Dreiner:1996dd,
RparityWorkingGroup:1999ixj,Datta:2000yc,Desai:2010sq,Bhattacherjee:2011dt,
Bhattacherjee:2013gr,Dercks:2017lfq,Domingo:2018qfg,
Bansal:2018dge,Dreiner:2020lbz,Barman:2020azo,
FileviezPerez:2022ypk,Dib:2022ppx,Choudhury:2023eje,
Choudhury:2023yfg,Choudhury:2024ggy,Domingo:2024qoj,
Choudhury:2024crp,Bickendorf:2024ovi,Baer:2025srs,Barman:2026cez}.
A first systematic
study of the collider phenomenology with a neutralino LSP was performed 
in Ref.~\cite{Dreiner:1991pe}. A further systematic study was performed 
in Ref.~\cite{Dercks:2017lfq}, however within the framework of the 
R-parity violating CMSSM. One R-parity violating coupling is assumed 
non-zero at the unification scale together with the CMSSM parameters: ($M_0,\, 
M_{1/2},\,\mu,\,A,\,\tan\beta$). The model is then evolved down to the 
weak scale. This generates the full supersymmetric spectrum as a function
of the six parameters; this includes all possible LSPs. The R-parity 
violating CMSSM is a specific constrained model. The full spectrum has 
the advantage of fixing all the relative supersymmetric masses and thus 
all the decay rates and branching ratios. The authors then checked for 
the coverage of the possible signatures at the LHC at the time. However 
this model is very specific and could be missing some signatures. 
Inspired by the work in Ref.~\cite{Konar:2010bi}, the authors in 
Ref.~\cite{Dreiner:2012wm} presented the most general set of R-parity 
violating signatures for any spectrum. This set of signatures is so 
general that it is not feasible to systematically search for all of them
at the LHC.

This paper is the third in a set of papers, that in turn investigate
the phenomenology with a dominant $LL\bar E$ \cite{Dreiner_2023}, a
dominant $\bar U\bar D\bar D$ coupling \cite{Dreiner:2025kfd}, and
now here a dominant $LQ\bar D$ coupling. The approach differs from
the R-parity violating CMSSM model investigated in 
Ref.~\cite{Dercks:2017lfq} in being more bottom-up. The model is only 
investigated at the weak scale; no renormalization group running is taken
into account. Thus, strictly only one coupling is considered at a time.
It is less general than in Ref.~\cite{Dreiner:2012wm} in that only direct
production and single cascade production are taken into account. In
Refs.~\cite{Dreiner_2023, Dreiner:2025kfd} the coverage of these models 
at the LHC was determined respectively. The purpose of this paper is to
complete the investigation of the superpotential in 
Eq.~\eqref{eq:RPV-superpot} and systematically investigate the coverage 
of the R-parity violating MSSM with a single $LQ\bar D$ coupling at the 
LHC.

The outline of this paper is as follows. In Sect.~\ref{sec:framework}
we briefly review the framework already applied in the previous two papers
\cite{Dreiner_2023, Dreiner:2025kfd}, and divide the 27 
$LQ\bar D$ couplings into benchmark classes. In Sect.~\ref{sec:LQD} we apply
this framework to classify the signatures for various LSPs decaying via the identified benchmark couplings, followed by a brief discussion on the relevant CMS and ATLAS analyses for the broad classifications of signatures. In Sect.~\ref{sec:results} we present our results for both the direct and cascade production of LSPs decaying via various $LQ\bar D$ couplings. Finally, we conclude in Sect.~\ref{sec:conclusion}.

\section{The ABC of RPV Framework in a Nutshell}
\label{sec:framework}

The ABC-RPV  
framework, first introduced in Ref.\,\cite{Dreiner_2023}, provides a 
systematic classification of RPV collider signatures in terms of the 
production mode, the LSP, and the relevant dominant RPV coupling. This 
classification offers a unified description of the wide range of final states 
that can arise in the RPV-MSSM at collider experiments such as the LHC. It 
thereby enables comprehensive studies of the current experimental coverage of 
these signatures, while identifying gaps that can be targeted by future 
searches. For simplicity, we restrict our analysis to spectra involving one
or two relevant sparticles at a time, rather than considering the general 
sparticle mass hierarchies explored in Ref.~\cite{Dreiner:2012wm}.

The ABC-RPV framework is currently limited to the regime where RPV 
interactions affect only the decay of the LSP, while SUSY production and 
cascade decays remain identical to those in the RPC-MSSM.  Consequently, SUSY particles are pair-produced through
gauge interactions and, unless the produced particle is itself the LSP, 
undergo the usual RPC cascade decays to the LSP. The LSP subsequently decays 
promptly via the $LQ\bar D$ interaction. 
For both the gauge-cascade production and the $LQ\bar D$ decay of the LSP, we consider the minimal number of decay steps to avoid additional decay-width suppression from heavy intermediate mediators. We generally restrict ourselves to one-step cascades, extending to two steps where necessary.
Throughout this work, all decays are therefore 
assumed to be prompt. For each benchmark scenario, the relevant $LQ\bar D$ 
coupling approximately satisfies\,\footnote{The precise range depends on the 
spectrum and on the specific $LQ\bar D$ coupling under consideration.}
\begin{eqnarray}
    \sqrt{\frac{(\beta\gamma)10^{-12} {\rm\,GeV}}{m_{\rm LSP}}}\lesssim 
    \lam' \ll g\,,
    \label{eq:range}
\end{eqnarray}
where $m_{\rm LSP}, \beta$ and $\gamma$ are, respectively, the mass, speed and
Lorentz boost factor of the LSP decaying via the $LQ\bar D$ coupling, $\lam'$,
and $g$ denotes a generic gauge coupling. The lower bound is obtained by 
requiring the LSP to have at most a decay length of approximately $1\, 
\mathrm{cm}$ in the laboratory frame, assuming a two-body decay. While the 
ABC-RPV framework can be extended to cover the long-lived LSP regime, a 
systematic and detailed investigation of this region of the parameter space 
lies beyond the scope of the present work and is left for future study.

For completeness and ease of reading, we first introduce the convention for 
the labelling of the various sparticles and SM particles appearing in 
the final states used in the present work in Table\,\ref{tab:conv}. As 
discussed in Sect.\,\ref{sec:intro}, the list of possible LSPs is much longer than when RPC is imposed. We will consider the following possible 
LSPs:
\begin{equation}
    \mathrm{LSP} \in \{\tilde{g},\tilde{q}_L,\tilde{q}_3,\tilde{u},\tilde{d},\tilde{t}_R,\tilde{b}_R,\widetilde{B},\widetilde{W},\widetilde{H},\tilde{\ell},\tilde{\nu},\tilde{\tau}_L,\tilde{\nu}_\tau,\tilde{\ell}_R,\tilde{\tau}_R\}\,.
\end{equation}
\begin{table}[hbt!]
    \centering
    \begin{tabular}{p{2.5cm} p{6cm}}
    \hline\hline
    Symbol     &  Particles\\
    \hline\hline
    $\ell$ & $e/\mu$ \\
    $L$ & $e/\mu/\tau$ \\
    $j_l$ & $u/d/c/s$ jets\\
    $j$ & $j_l$/$b$ jet/decay products of $t$ quark\\
    $V$     & $W/Z/h$\\
    $\tilde{q}_L$ & $\tilde{u}_L/\tilde{d}_L/\tilde{c}_L/\tilde{s}_L$\\
    $\tilde{q}_{3}$ & $\tilde{b}_L/\tilde{t}_L$ \\
    $\tilde{u}$ & $\tilde{u}_R/\tilde{c}_R$ \\
    $\tilde{d}$ & $\tilde{d}_R/\tilde{s}_R$ \\
    $\tilde{\ell}$ & $\tilde{e}_L/\tilde{\mu}_L$\\
    $\tilde{\nu}$ & $\tilde{\nu}_e/\tilde{\nu}_\mu$\\
    $\tilde{\ell}_R$ & $\tilde{e}_R/\tilde{\mu}_R$\\
    $\widetilde{W}$ & Winos ($\widetilde{W}^0$/$\widetilde{W}^\pm$)\\
    $\widetilde{H}$ & Higgsinos ($\widetilde{H}^0$/$\widetilde{H}^\pm$)\\
    \hline\hline
    \end{tabular}
    \caption{Convention for the various SM final states and sparticles used in the present work. For the particles not present in this list, we use the standard convention.}
    \label{tab:conv}
\end{table}
Each of these LSPs can either be produced directly at the collider, or via the
cascade decay of another sparticle, if the latter has a higher production 
cross section. 
The collider signature depends on the production mode, any subsequent cascade to the LSP, the identity of the LSP, and the $LQ\bar D$ operator through which it decays. Schematically, we write
\begin{eqnarray}
    {\rm Final\,state} \sim ({\rm Produced\,sparticle}) \otimes ({\rm LSP}) \otimes (LQ\bar D\,{\rm operator})\,.
\end{eqnarray}
Here $LQ\bar D$ refers to any of the $LQ\bar D$ couplings, $\lam'_{ijk}$,
in Eq.\,(\ref{eq:RPV-superpot}). The generation indices $i,j,k$ can take on
any value. Therefore, we have 27 independent $LQ\bar D$ couplings. Since the 
first two generations of light quarks are experimentally difficult to 
distinguish, and the reconstruction efficiencies for electrons and muons are similar, the set of $LQ\bar D$ couplings having unique collider signatures can be further reduced to eight representative couplings as shown in Table~\ref{tab:representative-couplings}.

\begin{table}
\begin{tabular}{l c l}\hline
Operators& Rep. $\lam'$ & Other Couplings \\ \hline
    \textbf{LQD} ($i,j,k \in \{1,2\}$) &  $\mathbf{\lambda'_{111}}$ & $\lambda'_{112}$, $\lam'_{121}$, $\lam'_{122}$, $\lambda'_{211}$, $\lambda'_{212}$, $\lam'_{221}$, $\lambda'_{222}$\\
    {\bf LQD$_3$} ($i,j \in \{1,2\}$, $k=3$) & $\mathbf{\lambda'_{113}}$ & $\lambda'_{123}$, $\lam'_{213}$, $\lambda'_{223}$\\
    {\bf LQ$_3$D} ($i,k \in \{1,2\}$, $j=3$) & $\mathbf{\lambda'_{131}}$ & $\lambda'_{132}$, $\lam'_{231}$, $\lam'_{232}$\\
    {\bf LQ$_3$D$_3$} ($i \in \{1,2\}$, $j=k=3$) & $\mathbf{\lam'_{133}}$ & $\lambda'_{233}$\\
    {\bf L$_3$QD} ($i=3$, $j,k \in \{1,2\}$) & $\mathbf{\lambda'_{311}}$ & $\lambda'_{312}$, $\lambda'_{321}$, $\lambda'_{322}$\\
    {\bf L$_3$QD$_3$} ($i=k=3$, $j \in \{1,2\}$) & $\mathbf{\lambda'_{313}}$ & $\lambda'_{323}$\\
    {\bf L$_3$Q$_3$D} ($i=j=3$, $k \in \{1,2\}$) & $\mathbf{\lambda'_{331}}$ & $\lambda'_{332}$\\
    {\bf L$_3$Q$_3$D$_3$} ($i=j=k=3$) & $\mathbf{\lam'_{333}}$ & \\\hline
\end{tabular}
\caption{Classification of the $LQ\bar D$ couplings into eight classes with distinct collider signatures. For each class, the second column shows the representative coupling used in our numerical analysis.}
\label{tab:representative-couplings}
\end{table}

In the next section we present the final states obtained for LSPs decaying via
$LQ\bar D$ couplings. For each LSP, we consider the relevant production modes 
at the LHC, either directly or via a cascade, and study its decay via the eight
benchmark $LQ\bar D$ couplings. 
We assume that only the LSP and the produced sparticles are kinematically accessible at the LHC, while all other sparticles are decoupled.

%
\section{Application of the Framework to LQD Couplings}
\label{sec:LQD}

Here we discuss the benchmark scenarios for each possible LSP and the 
corresponding LHC final states. Based on these final states, we then identify
the relevant CMS and ATLAS analyses and assess their availability within the 
recasting framework \texttt{CheckMATE\,2}.

\subsection{Benchmark Scenarios}
\label{ssec:benchmarks}

In the following, we discuss in turn the possible production modes and 
final states for each possible LSP decaying via the benchmark $LQ\bar D$ 
couplings. Depending on the identity of the LSP and the dominant $LQ\bar D$ 
coupling, the LSP may decay either directly into SM particles or via one or 
more intermediate off-shell sparticles, with the $LQ\bar D$ interaction 
occurring in the final step. Throughout this section, we denote direct 
(zero-step) decays in {\it \textcolor{greeen}{blue}}, one-step decays in {\it 
\textcolor{newpurple}{purple}}, and two-step decays in {\it black}.

\subsubsection*{Gluino LSP}

Owing to their potentially large production cross section at the LHC, we 
consider the direct pair production of gluino LSPs and their subsequent decays 
for each of the eight benchmark $LQ\bar D$ couplings of 
Table~\ref{tab:representative-couplings}. We keep all the other sparticles 
decoupled from the considered spectrum, particularly beyond the kinematic
reach of the LHC. The gluino decays via an 
off-shell squark, $$\tilde g \to j_l + (\tilde q)^\star \to 2j_l + (\nu/l) \,,$$ 
where $(\cdots)^\star$ denotes an off-shell intermediate sparticle. 
The $LQ\bar D$ interaction occurs at the final vertex, in the 
decay of the virtual squark. Each gluino decay therefore produces two quarks 
together with either a charged lepton or a neutrino. The flavours of the 
final-state quarks and leptons are determined by the dominant $LQ\bar D$ 
coupling. The corresponding decay chains can be obtained using the RPV Python 
library \texttt{abc-rpv}~\cite{Dreiner_2023,abc-rpv}. The resulting final 
states for the eight benchmark $LQ\bar D$ couplings are summarised in 
Table~\ref{tab:gluino_LQD}. Due to the $SU(2)_L$ structure of the $LQ\bar D$ 
operator, each coupling gives rise to two possible decay topologies, involving 
either a charged lepton and an up-type quark or a neutrino\,\footnote{The
neutrino gives rise to missing transverse energy (MET) in collider detectors at
the LHC.} and a down-type quark.

\begin{table}[hbt!]
    \centering
    \resizebox{0.7\textwidth}{!}{
    \begin{tabular}{c c c}
    \hline\hline
    LSP  &   Coupling  &  LSP Decay \\
    \hline\hline
    \multirow{8}{*}{$\widetilde{g}$} &  $\lambda_{111}'$ &\textcolor{newpurple}{$2j_l+1l$~~~~~~~~$2j_l+$MET}            \\
                                   &  $\lambda_{113}'$ & \textcolor{newpurple}{$1j_l+1b+1l$ ~~~~~~~~$1j_l+1b+$MET}          \\
                                   &  $\lambda_{131}'$ & \textcolor{newpurple}{$1j_l+1t+1l$ ~~~~~~~~$1j_l+1b+$MET}                       \\
                                   &  $\lambda_{133}'$ & \textcolor{newpurple}{$1t+1b+1l$ ~~~~~~~~$2b+$MET}                               \\
                                   &  $\lambda_{311}'$ & \textcolor{newpurple}{$2j_l+1\tau$ ~~~~~~~~$2j_l+$MET}                   \\
                                   &  $\lambda_{313}'$ & \textcolor{newpurple}{$1j_l+1b+1\tau$ ~~~~~~~~$1j_l+1b+$MET}        \\
                                   &  $\lambda_{331}'$ & \textcolor{newpurple}{$1j_l+1t+1\tau$ ~~~~~~~~$1j_l+1b+$MET}                         \\
                                   &  $\lambda_{333}'$ & \textcolor{newpurple}{$1t+1b+1\tau$ ~~~~~~~~$2b+$MET}                            \\
                                     
    \hline\hline  
    \end{tabular}
    }
    \caption{Details of the gluino LSP benchmarks: the first column 
    depicts the LSP; the $LQ\bar D$ coupling assumed to be non-zero is shown in
    the second column; and the last column represents the final states
    from the individual LSP decay.
We show the final states from the one-step cascade decay in {\it \textcolor{newpurple}{purple}}.}
    \label{tab:gluino_LQD}    
\end{table}

\begin{table}[hb!]
    \centering
    \resizebox{0.9\linewidth}{!}{
    \begin{tabular}{c | c c c c c c c c}
    \hline\hline
    \multirow{2}{*}{LSP} & \multicolumn{8}{c}{Coupling} \\ \cline{2-9}
         & $\lambda'_{111}$ & $\lambda'_{113}$ & $\lambda'_{131}$ & $\lambda'_{133}$ & $\lambda'_{311}$ & $\lambda'_{313}$ & $\lambda'_{331}$ & $\lambda'_{333}$ \\
    \hline\hline
    $\widetilde{u}_R/\widetilde{t}_R$ & \multicolumn{8}{c}{$\longleftarrow$ Cascade $\longrightarrow$} \\
    $\widetilde{d}_R$ &  \textcolor{greeen}{Direct} &  Cascade & \textcolor{greeen}{Direct} &  Cascade & \textcolor{greeen}{Direct} & Cascade & \textcolor{greeen}{Direct} & Cascade \\
    $\widetilde{b}_R$ & Cascade &  \textcolor{greeen}{Direct} & Cascade &  \textcolor{greeen}{Direct} & Cascade &  \textcolor{greeen}{Direct} & Cascade &  \textcolor{greeen}{Direct} \\
    $\widetilde{q}_L$ & \textcolor{greeen}{Direct} & \textcolor{greeen}{Direct} & Cascade & Cascade & \textcolor{greeen}{Direct} & \textcolor{greeen}{Direct} & Cascade & Cascade \\
    $\widetilde{t}_L/\widetilde{b}_L$ & Cascade & Cascade & \textcolor{greeen}{Direct} & \textcolor{greeen}{Direct} & Cascade & Cascade & \textcolor{greeen}{Direct} & \textcolor{greeen}{Direct} \\
    \hline\hline
    \end{tabular}
    }
    \caption{Direct or cascade decays of the various squark LSPs for the eight benchmark $LQ\bar D$ couplings of Table~\ref{tab:representative-couplings}.}
    \label{tab:squark_decays}
\end{table}

\subsubsection*{Squark LSPs}

For squark LSPs, we also consider their direct pair production at the LHC. 
In Table~\ref{tab:squark_decays}, we indicate, for each squark LSP and each of 
the eight benchmark $LQ\bar D$ couplings of 
Table~\ref{tab:representative-couplings}, whether the decay proceeds 
``directly'' (zero-step) or via a ``cascade''. The cascade decays for squarks 
always proceed in two steps, where the first step is mediated by an off-shell 
gluino, bino, wino, or even higgsino for the third generation, while the second
step involves an off-shell squark or slepton which can take part in the $LQ\bar
D$ interaction depending on the coupling. Tables~\ref{tab:RHsquark_LQD} and 
\ref{tab:LHsquark_LQD} show the final states for the different squark LSPs for 
each of the identified benchmark couplings. 

\begin{table}[hbt!]
    \centering
    \resizebox{0.8\textwidth}{!}{
    \begin{tabular}{c c c}
    \hline\hline
    LSP  &   Coupling  &  LSP Decay  \\
    \hline\hline
    \multirow{8}{*}{$\widetilde{u}_R$} &  $\lambda_{111}'$ & $3j_l+1l$~~~~~~~~$3j_l+$MET              \\
                                   &  $\lambda_{113}'$ & $2j_l+1b+1l$~~~~~~~~$2j_l+1b+$MET          \\
                                   &  $\lambda_{131}'$ & $2j_l+1t+1l$~~~~~~~~$2j_l+1b+$MET                             \\
                                   &  $\lambda_{133}'$ & $1j_l+1t+1b+1l$~~~~~~~~$1j_l+2b+$MET                             \\
                                   &  $\lambda_{311}'$ &       $3j_l+1\tau$~~~~~~~~$3j_l+$MET                         \\
                                   &  $\lambda_{313}'$ &       $2j_l+1b+1\tau$~~~~~~~~$2j_l+1b+$MET                        \\
                                   &  $\lambda_{331}'$ & $2j_l+1t+1\tau$~~~~~~~~$2j_l+1b+$MET                               \\
                                   &  $\lambda_{333}'$ & $1j_l+1t+1b+1\tau$~~~~~~~~$1j_l+2b+$MET                               \\
                                   \hline
    \multirow{8}{*}{$\widetilde{d}_R$} &  $\lambda_{111}'$ & \textcolor{greeen}{$1j_l+1l$~~~~~~~~$1j_l+$MET}              \\
                                   &  $\lambda_{113}'$ & $2j_l+1b+1l$~~~~~~~~$2j_l+1b+$MET          \\
                                   &  $\lambda_{131}'$ & \textcolor{greeen}{$1t+1l$~~~~~~~~$1b+$MET}                              \\
                                   &  $\lambda_{133}'$ & $1j_l+1t+1b+1l$~~~~~~~~$1j_l+2b+$MET                            \\
                                   &  $\lambda_{311}'$ &       \textcolor{greeen}{$1j_l+1\tau$~~~~~~~~$1j_l+$MET}                       \\
                                   &  $\lambda_{313}'$ &       $2j_l+1b+1\tau$~~~~~~~~$2j_l+1b+$MET                        \\
                                   &  $\lambda_{331}'$ & \textcolor{greeen}{$1t+1\tau$~~~~~~~~$1b+$MET}                            \\
                                   &  $\lambda_{333}'$ & $1j_l+1t+1b+1\tau$~~~~~~~~$1j_l+2b+$MET                               \\
                                   \hline
    \multirow{8}{*}{$\widetilde{t}_R$} &  $\lambda_{111}'$ & $2j_l+1t+1l$~~~~~~~~$2j_l+1t+$MET    \\
                                   &  $\lambda_{113}'$ & $1j_l+1t+1b+1l$~~~~~~~~$1j_l+1t+1b+$MET            \\
                                   &  $\lambda_{131}'$ & $1j_l+2t+1l$~~~~~~~~$1j_l+1t+1b+$MET~~~~~~~~$1j_l+2b+1l$\,$^{(*)}$                                 \\
                                   &  $\lambda_{133}'$ & $2t+1b+1l$~~~~~~~~$1t+2b+$MET~~~~~~~~$3b+1l$\,$^{(*)}$                              \\
                                   &  $\lambda_{311}'$ &       $2j_l+1t+1\tau$~~~~~~~~$2j_l+1t+$MET                       \\
                                   &  $\lambda_{313}'$ &       $1j_l+1t+1b+1\tau$~~~~~~~~$1j_l+1t+1b+$MET                       \\
                                   &  $\lambda_{331}'$ & $1j_l+2t+1\tau$~~~~~~~~$1j_l+1t+1b+$MET~~~~~~~~$1j_l+2b+1\tau$\,$^{(*)}$                               \\
                                   &  $\lambda_{333}'$ & $2t+1b+1\tau$~~~~~~~~$1t+2b+$MET~~~~~~~~$3b+1\tau$\,$^{(*)}$                               \\
                                   \hline    
    \multirow{8}{*}{$\widetilde{b}_R$} &  $\lambda_{111}'$ & $2j_l+1b+1l$~~~~~~~~$2j_l+1b+$MET            \\
                                   &  $\lambda_{113}'$ & \textcolor{greeen}{$1j_l+1l$~~~~~~~~$1j_l+$MET}            \\
                                   &  $\lambda_{131}'$ & $1j_l+1b+1t+1l$~~~~~~~~$1j_l+2b+$MET~~~~~~~~$1j_l+2t+$MET\,$^{(*)}$                                 \\
                                   &  $\lambda_{133}'$ & \textcolor{greeen}{$1t+1l$~~~~~~~~$1b+$MET}                               \\
                                   &  $\lambda_{311}'$ &       $2j_l+1b+1\tau$~~~~~~~~$2j_l+1b+$MET                        \\
                                   &  $\lambda_{313}'$ &       \textcolor{greeen}{$1j_l+1\tau$~~~~~~~~$1j_l+$MET}                      \\
                                   &  $\lambda_{331}'$ & $1j_l+1t+1b+1\tau$~~~~~~~~$1j_l+2b+$MET~~~~~~~~$1j_l+2t+$MET\,$^{(*)}$                              \\
                                   &  $\lambda_{333}'$ & \textcolor{greeen}{$1t+1\tau$~~~~~~~~$1b+$MET}                              \\    
    \hline\hline                            
    \end{tabular}
    }
    \caption{Details of the right-handed squark LSP benchmarks with columns as in Table\,\,\ref{tab:gluino_LQD}.
    We show the final states from the direct decay (two-step cascade decay) in {\it \textcolor{greeen}{blue}} ({\it black}). The decay modes marked by $^{(*)}$ are mediated exclusively by charged higgsinos ($\widetilde{H}^\pm$).}
    \label{tab:RHsquark_LQD}    
\end{table}

The final states arising from a $\tilde{d}_R$ LSP decaying via the $\lam'_ 
{ijk}$ couplings with $k=1,2$ are identical to those obtained for a $\tilde{b} 
_R$ LSP with couplings corresponding to $k=3$. A similar correspondence exists 
for the left-handed squark LSPs, where the decays of $\tilde{u}_L$ ($\tilde{d} 
_L$) via $\lam'_{i1k}$ lead to the same final states as the decays of $\tilde 
{t}_L$ ($\tilde{b}_L$) via $\lam'_{i3k}$. However, for left-handed squark LSPs,
only one of the two possible $SU(2)_L$ combinations in the $LQ\bar D$ operator 
contributes, resulting in only a single possible final state. If the 
$\tilde{u}_L$ and $\tilde{d}_L$ are degenerate in mass, they are produced with 
equal cross sections. Consequently, the inclusive signal contains equal 
fractions of the final states arising from their respective decays.

\begin{table}[hbt!]
    \centering
    \resizebox{0.9\textwidth}{!}{
    \begin{tabular}{c c c}
    \hline\hline
    LSP  &   Coupling  &  LSP Decay  \\
    \hline\hline
    \multirow{8}{*}{$\widetilde{q}_L$} &  $\lambda_{111}'$ & \textcolor{greeen}{$1j_l+1l$~~~~~~~~$1j_l+$MET}             \\
                                   &  $\lambda_{113}'$ & \textcolor{greeen}{$1b+1l$~~~~~~~~$1b+$MET}           \\
                                   &  $\lambda_{131}'$ & $2j_l+1t+1l$~~~~~~~~$2j_l+1t+$MET\,$^{(**)}$~~~~~~~~$2j_l+1b+1l$\,$^{(**)}$~~~~~~~~$2j_l+1b+$MET                          \\
                                   &  $\lambda_{133}'$ & $1j_l+1t+1b+1l$~~~~~~~~$1j_l+1t+1b+$MET\,$^{(**)}$~~~~~~~~$1j_l+2b+1l$\,$^{(**)}$~~~~~~~~$1j_l+2b+$MET                      \\
                                   &  $\lambda_{311}'$ &       \textcolor{greeen}{$1j_l+1\tau$~~~~~~~~$1j_l+$MET}                   \\
                                   &  $\lambda_{313}'$ &       \textcolor{greeen}{$1b+1\tau$~~~~~~~~$1b+$MET}                   \\
                                   &  $\lambda_{331}'$ & $2j_l+1t+1\tau$~~~~~~~~$2j_l+1t+$MET\,$^{(**)}$~~~~~~~~$2j_l+1b+1\tau$\,$^{(**)}$~~~~~~~~$2j_l+1b+$MET                        \\
                                   &  $\lambda_{333}'$ & $1j_l+1t+1b+1\tau$~~~~~~~~$1j_l+1t+1b+$MET\,$^{(**)}$~~~~~~~~$1j_l+2b+1\tau$\,$^{(**)}$~~~~~~~~$1j_l+2b+$MET                      \\
                                   \hline
    \multirow{8}{*}{$\widetilde{t}_L$} &  $\lambda_{111}'$ & $2j_l+1t+1l$~~~~~~~~$2j_l+1t+$MET~~~~~~~~$2j_l+1b+1l$\,$^{(**)}$~~~~~~~~$2j_l+1b+$MET\,$^{(**)}$     \\
                                   &  $\lambda_{113}'$ & $1j_l+1t+1b+1l$~~~~~~~~$1j_l+1t+1b+$MET~~~~~~~~$1j_l+2b+1l$\,$^{(**)}$~~~~~~~~$1j_l+2b+$MET\,$^{(**)}$            \\
                                   &  $\lambda_{131}'$ & \textcolor{greeen}{$1j_l+1l$}                              \\
                                   &  $\lambda_{133}'$ & \textcolor{greeen}{$1b+1l$}                               \\
                                   &  $\lambda_{311}'$ &       $2j_l+1t+1\tau$~~~~~~~~$2j_l+1t+$MET~~~~~~~~$2j_l+1b+1\tau$\,$^{(**)}$~~~~~~~~$2j_l+1b+$MET\,$^{(**)}$                     \\
                                   &  $\lambda_{313}'$ &       $1j_l+1t+1b+1\tau$~~~~~~~~$1j_l+1t+1b+$MET~~~~~~~~$1j_l+2b+1\tau$\,$^{(**)}$~~~~~~~~$1j_l+2b+$MET\,$^{(**)}$                       \\
                                   &  $\lambda_{331}'$ & \textcolor{greeen}{$1j_l+1\tau$}                             \\
                                   &  $\lambda_{333}'$ & \textcolor{greeen}{$1b+1\tau$}                          \\
                                   \hline    
    \multirow{8}{*}{$\widetilde{b}_L$} &  $\lambda_{111}'$ & $2j_l+1b+1l$~~~~~~~~$2j_l+1b+$MET~~~~~~~~$2j_l+1t+1l$\,$^{(**)}$~~~~~~~~$2j_l+1t+$MET\,$^{(**)}$         \\
                                   &  $\lambda_{113}'$ & $1j_l+2b+1l$~~~~~~~~$1j_l+2b+$MET~~~~~~~~$1j_l+1t+1b+1l$\,$^{(**)}$ \\
                                   & & $1j_l+1t+1b+$MET\,$^{(**)}$~~~~~~~~$1j_l+2t+1l$\,$^{(*)}$~~~~~~~~$1j_l+2t+$MET\,$^{(*)}$              \\
                                   &  $\lambda_{131}'$ & \textcolor{greeen}{$1j_l+$MET}                                 \\
                                   &  $\lambda_{133}'$ & \textcolor{greeen}{$1b+$MET}                               \\
                                   &  $\lambda_{311}'$ &       $2j_l+1b+1\tau$~~~~~~~~$2j_l+1b+$MET~~~~~~~~ $2j_l+1t+1\tau$\,$^{(**)}$~~~~~~~~$2j_l+1t+$MET\,$^{(**)}$                      \\
                                   &  $\lambda_{313}'$ &    $1j_l+2b+1\tau$~~~~~~~~$1j_l+2b+$MET~~~~~~~~$1j_l+1t+1b+1\tau$\,$^{(**)}$  \\
                                   & & $1j_l+1t+1b+$MET\,$^{(**)}$~~~~~~~~$1j_l+2t+1\tau$\,$^{(*)}$~~~~~~~~$1j_l+2t+$MET\,$^{(*)}$               \\
                                   &  $\lambda_{331}'$ & \textcolor{greeen}{$1j_l+$MET}                            \\
                                   &  $\lambda_{333}'$ & \textcolor{greeen}{$1b+$MET}                           \\   
    \hline\hline                            
    \end{tabular}
    }
    \caption{Details of the left-handed squark LSP benchmarks with columns as in Table\,\,\ref{tab:gluino_LQD}.
    We show the final states from the direct decay (two-step cascade decay) in {\it \textcolor{greeen}{blue}} ({\it black}). The decay modes marked by $^{(*)}$ and $^{(**)}$ are mediated exclusively by charged higgsinos ($\widetilde{H}^\pm$) and charged winos ($\widetilde{W}^\pm$), respectively.}
    \label{tab:LHsquark_LQD}    
\end{table}

In cases where more than two final states are possible, the additional 
signatures arise from the different intermediate off-shell sparticles that can 
mediate the decay. A representative example is the $\tilde{b}_R$ ($\tilde{t}_R 
$) LSP decaying via the $\lam'_{131}$ and $\lam'_{331}$ couplings, 
respectively. Besides the final states obtained through neutral intermediate 
states, an additional final state, $1j_l+2t+\mathrm{MET}$ ($1j_l+2b+1l/1\tau$),
can be produced through an intermediate off-shell charged higgsino. The 
decay chain for $\tilde{b}_R$ for this final state is
$$\tilde{b}_R \rightarrow t + (\widetilde{H}^+)^\star
\rightarrow 2t + (\tilde{b}_L)^\star
\rightarrow 2t + \nu_e + j_l\,,$$
where $(\cdots)^\star$ denotes an off-shell intermediate sparticle.
A similar situation arises for the $\tilde{b}_L$ LSP, where decays mediated by the charged wino and charged higgsino give rise to final states that differ from those obtained through the neutral wino, neutral higgsino, bino, or gluino. The relative branching fractions among these final states are determined by the next-to-lightest supersymmetric particle (NLSP) and the corresponding SM gauge and Yukawa couplings.


\subsubsection*{Electroweakino LSPs}

In the electroweakino sector, we first consider the possibility 
of a bino-like LSP, $\widetilde{B}$, with the rest of the 
sparticles decoupled from the spectrum. Table~\ref{tab:bino_LQD} 
shows the final states for the bino-like LSP for each of the 
eight $LQ\bar D$ benchmark couplings \cite{Dreiner:1991pe}. The 
decay modes are the same as the gluino LSP, and it proceeds via a
one-step cascade through a slepton, sneutrino or squark off-shell
mediator. Since the direct pair production cross section of pure 
bino-like LSPs at the LHC is very small, we 
consider only its production through cascade decays of heavier 
sparticles, such as gluinos, squarks, sleptons, and wino-like 
electroweakinos. 
Depending on the production mode, the cascade gives rise to 
additional jets, leptons, vector bosons, or Higgs bosons in the 
final state, accompanying the bino LSP.

\begin{table}[t!]
    \centering
    \resizebox{0.7\textwidth}{!}{
    \begin{tabular}{c c c}
    \hline\hline
    LSP  &   Coupling  &  LSP Decay   \\
    \hline\hline
    \multirow{8}{*}{$\widetilde{B}$} &  $\lambda_{111}'$ & \textcolor{newpurple}{$2j_l+1l$~~~~~~~~$2j_l+$MET}              \\
                                   &  $\lambda_{113}'$ & \textcolor{newpurple}{$1j_l+1b+1l$ ~~~~~~~~$1j_l+1b+$MET}              \\
                                   &  $\lambda_{131}'$ & \textcolor{newpurple}{$1j_l+1t+1l$ ~~~~~~~~$1j_l+1b+$MET}             \\
                                   &  $\lambda_{133}'$ & \textcolor{newpurple}{$1t+1b+1l$ ~~~~~~~~$2b+$MET}                    \\
                                   &  $\lambda_{311}'$ & \textcolor{newpurple}{$2j_l+1\tau$ ~~~~~~~~$2j_l+$MET}                   \\
                                   &  $\lambda_{313}'$ & \textcolor{newpurple}{$1j_l+1b+1\tau$ ~~~~~~~~$1j_l+1b+$MET}            \\
                                   &  $\lambda_{331}'$ & \textcolor{newpurple}{$1j_l+1t+1\tau$ ~~~~~~~~$1j_l+1b+$MET}                           \\
                                   &  $\lambda_{333}'$ & \textcolor{newpurple}{$1t+1b+1\tau$ ~~~~~~~~$2b+$MET}                         \\
                                     
    \hline\hline  
    \end{tabular}
    }
    \caption{Details of the bino LSP benchmarks with columns as in Table\,\,\ref{tab:gluino_LQD}. We show the final states from the one-step cascade decay in {\it \textcolor{newpurple}{purple}}.}
    \label{tab:bino_LQD}    
\end{table}

\begin{table}[hbt!]
    \centering
    \resizebox{0.7\textwidth}{!}{
    \begin{tabular}{c c c}
    \hline\hline
    LSP  &   Coupling  &  LSP Decay   \\
    \hline\hline
    \multirow{8}{*}{$\widetilde{W}^+$} &  $\lambda_{111}'$ &\textcolor{newpurple}{$2j_l+$MET~~~~~~~~$2j_l+1l$}             \\
                                   &  $\lambda_{113}'$ & \textcolor{newpurple}{$1j_l+1b+1l$ ~~~~~~~~$1j_l+1b+$MET}             \\
                                   &  $\lambda_{131}'$ & \textcolor{newpurple}{$1j_l+1t+$MET ~~~~~~~~$1j_l+1b+1l$ }                      \\
                                   &  $\lambda_{133}'$ & \textcolor{newpurple}{$1t+1b+$MET ~~~~~~~~$2b+1l$}                              \\
                                   &  $\lambda_{311}'$ &\textcolor{newpurple}{$2j_l+1\tau$ ~~~~~~~~$2j_l+$MET}                    \\
                                   &  $\lambda_{313}'$ &\textcolor{newpurple}{$1j_l+1b+1\tau$ ~~~~~~~~$1j_l+1b+$MET}            \\
                                   &  $\lambda_{331}'$ &\textcolor{newpurple}{ $1j_l+1t+$MET ~~~~~~~~$1j_l+1b+1\tau$}                          \\
                                   &  $\lambda_{333}'$ &\textcolor{newpurple}{$1t+1b+$MET ~~~~~~~~$2b+1\tau$}                          \\
                                     
    \hline  
    \multirow{8}{*}{$\widetilde{W}^0$} &  $\lambda_{111}'$ &\textcolor{newpurple}{$2j_l+$MET~~~~~~~~$2j_l+1l$}\\
    &$\lambda_{113}'$&\textcolor{newpurple}{$1j_l+1b+1l$ ~~~~~~~~$1j_l+1b+$MET} \\
    
    &$\lambda_{131}'$&\textcolor{newpurple}{$1j_l+1t+1l$ ~~~~~~~~$1j_l+1b+$MET} \\
     &$\lambda_{133}'$&\textcolor{newpurple}{$1t+1b+1l$ ~~~~~~~~$2b+$MET} \\
    &$\lambda_{311}'$&\textcolor{newpurple}{$2j_l+1\tau$ ~~~~~~~~$2j_l+$MET} \\
    &$\lambda_{313}'$&\textcolor{newpurple}{$1j_l+1b+1\tau$ ~~~~~~~~$1j_l+1b+$MET}\\
    &$\lambda_{331}'$& \textcolor{newpurple}{$1j_l+1t+1\tau$ ~~~~~~~~$1j_l+1b+$MET}\\
    &$\lambda_{333}'$&\textcolor{newpurple}{$1t+1b+1\tau$ ~~~~~~~~$2b+$MET}\\
    \hline\hline 
    \end{tabular}
    }
    \caption{Details of the wino LSP benchmarks with columns as in Table\,\,\ref{tab:gluino_LQD}.
    We show the final states from the one-step decay in {\it \textcolor{newpurple}{purple}}. Two- and higher-step decays are not shown.}
    \label{tab:Wino_LQD}    
\end{table}

\begin{table}[hbt!]
    \centering
    \resizebox{0.7\textwidth}{!}{
    \begin{tabular}{c c c}
    \hline\hline
    LSP  &   Coupling  &  LSP Decay \\
    \hline\hline
    \multirow{8}{*}{$\widetilde{H}^+$} &  $\lambda_{111}'$ &$2j_l+1V+$MET~~~~~~~~$2j_l+1V+1l$             \\
                                   &  $\lambda_{113}'$ & \textcolor{newpurple}{$1j_l+1t+1l$~~~~~~~~$1j_l+1t+$MET}             \\
                                   &  $\lambda_{131}'$ & \textcolor{newpurple}{$1j_l+1t+$MET ~~~~~~~~$1j_l+1b+1l$ }                  \\
                                   &  $\lambda_{133}'$ & \textcolor{newpurple}{$1t+1b+$MET ~~~~~~~~$2b+1l$}                                 \\
                                   &  $\lambda_{311}'$ &{$2j_l+1\tau+1V$ ~~~~~~~~$2j_l+1V+$MET}                     \\
                                   &  $\lambda_{313}'$ &\textcolor{newpurple}{$1j_l+1t+1\tau$ ~~~~~~~~$1j_l+1t+$MET}             \\
                                   &  $\lambda_{331}'$ &\textcolor{newpurple}{ $1j_l+1t+$MET ~~~~~~~~$1j_l+1b+1\tau$}                          \\
                                   &  $\lambda_{333}'$ &\textcolor{newpurple}{$1t+1b+$MET ~~~~~~~~$2b+1\tau$~~~~~~~~$2t+1\tau$}                            \\
                                     
    \hline  
    \multirow{8}{*}{$\widetilde{H}^0$} &  $\lambda_{111}'$ &$2j_l+1V+$MET~~~~~~~~$2j_l+1V+1l$\\
    &$\lambda_{113}'$&\textcolor{newpurple}{$1j_l+1b+1l$ ~~~~~~~~$1j_l+1b+$MET} \\
    
    &$\lambda_{131}'$&\textcolor{newpurple}{$1j_l+1t+1l$ ~~~~~~~~$1j_l+1b+$MET} \\
     &$\lambda_{133}'$&\textcolor{newpurple}{$1t+1b+1l$ ~~~~~~~~$2b+$MET} \\
    &$\lambda_{311}'$&{$2j_l+1\tau+1V$ ~~~~~~~~$2j_l+1V+$MET}   \\
    &$\lambda_{313}'$&\textcolor{newpurple}{$1j_l+1b+1\tau$ ~~~~~~~~$1j_l+1b+$MET}\\
    &$\lambda_{331}'$& \textcolor{newpurple}{$1j_l+1t+1\tau$ ~~~~~~~~$1j_l+1b+$MET}\\
    &$\lambda_{333}'$&\textcolor{newpurple}{$1t+1b+1\tau$ ~~~~~~~~$2b+$MET}\\
    \hline\hline 
    \end{tabular}
    }
    \caption{Details of the higgsino LSP benchmarks with columns as in Table\,\,\ref{tab:gluino_LQD}.
    We show the final states from the one-step decay (two-step decay) in {\it \textcolor{newpurple}{purple}} (\textit{black}).}
    \label{tab:Higgsino_LQD}    
\end{table}

The wino-like LSPs constitute a neutralino ($\tilde{\chi}_1^0$) 
and a pair of charginos ($\tilde{\chi}_1^\pm$), all having 
nearly-degenerate masses. The dominant direct production channels
of wino-like LSPs are $pp\to\tilde{\chi}_1^\pm\tilde{\chi}_1^0$ 
and $pp\to\tilde{\chi}_1^+\tilde{\chi}_1^-$. 
Table~\ref{tab:Wino_LQD} shows the final states for the wino-like
neutralino and chargino LSPs for each of the identified benchmark
couplings. The decay modes of a wino-like neutralino 
($\widetilde{W}^0$) are the same as those of a bino-like 
neutralino, which proceed via a one-step cascade. The decay modes
of wino-like charginos ($\widetilde{W}^\pm$) are also the same 
for all the couplings, except for $\lambda'_{i3k}$, $i,k\in
\{1,3\}$.

For higgsino-like LSPs, the spectrum now includes degenerate neutralinos ($\tilde{\chi}_1^0$, $\tilde{\chi}_2^0$) and a pair of charginos ($\tilde{\chi}_1^\pm$).
The dominant direct production modes at the LHC are $pp\to\tilde{\chi}_1^\pm\tilde{\chi}_{1/2}^0$, $pp\to\tilde{\chi}_1^+\tilde{\chi}_1^-$ and $pp\to\tilde{\chi}_1^0\tilde{\chi}_2^0$.
Table~\ref{tab:Higgsino_LQD} shows the final states for the 
higgsino-like neutralino and chargino LSPs for each of the 
identified benchmark couplings. The decay topologies of the 
charged and neutral higgsino-like LSPs ($\widetilde{H}^{\pm}$, 
$\widetilde{H}^0$) are largely identical to those of the 
corresponding wino-like LSPs. This is because, for couplings 
involving at least one third-generation quark superfield, the 
large top and bottom Yukawa couplings allow the higgsino-like LSP
to decay through a one-step cascade, analogous to the wino case. 
An exception arises for the couplings involving only the first 
two generations of quark superfields, $\lam'_{i11},\,\lam'_ 
{i12},\,\lam'_{i21},\,\mathrm{and}\;\lam'_{i22}$, $i\in\{1,2 ,3 
\}$. In this case, the small Yukawa couplings suppress the 
one-step decay, and the higgsino-like LSP instead undergoes a 
two-step cascade. For example,
$$\tilde{H}^0 \rightarrow V + (\widetilde{B})^\star
\rightarrow j_l + (\tilde{d}_R)^\star
\rightarrow V + 2j_l + (e/\nu_e)\,,$$
where $V=W,Z,h$. Consequently, the final state contains vector bosons or Higgs bosons compared to the corresponding decay modes of wino and bino-like LSPs.

\subsubsection*{Slepton and Sneutrino LSPs}

Similar to the squark LSPs, the structure of the $LQ\bar D$ couplings allows some slepton and sneutrino LSPs to decay directly, depending on the dominant coupling.
For each slepton and sneutrino LSP, we indicate whether the decay proceeds directly or via a cascade for each of the eight benchmark $LQ\bar D$ couplings in Table~\ref{tab:slepton_decays}. 

\begin{table}[hbt!]
    \centering
    \resizebox{0.9\linewidth}{!}{
    \begin{tabular}{c | c c c c c c c c}
    \hline\hline
    \multirow{2}{*}{LSP} & \multicolumn{8}{c}{Coupling} \\ \cline{2-9}
         & $\lambda'_{111}$ & $\lambda'_{113}$ & $\lambda'_{131}$ & $\lambda'_{133}$ & $\lambda'_{311}$ & $\lambda'_{313}$ & $\lambda'_{331}$ & $\lambda'_{333}$ \\
    \hline\hline
    $\widetilde{e}_R/\widetilde{\mu}_R/\widetilde{\tau}_R$ & \multicolumn{8}{c}{$\longleftarrow$ Cascade $\longrightarrow$} \\
    $\widetilde{e}_L/\widetilde{\mu}_L/\widetilde{\nu}_e/\widetilde{\nu}_\mu$ &  \textcolor{greeen}{Direct} &   \textcolor{greeen}{Direct} &   \textcolor{greeen}{Direct} &  \textcolor{greeen}{Direct} & Cascade & Cascade & Cascade & Cascade \\
    $\widetilde{\tau}_L/\widetilde{\nu}_\tau$ & Cascade & Cascade & Cascade & Cascade & \textcolor{greeen}{Direct} &   \textcolor{greeen}{Direct} &   \textcolor{greeen}{Direct} &  \textcolor{greeen}{Direct} \\
    \hline\hline
    \end{tabular}
    }
    \caption{Direct or cascade decays of the various slepton LSPs for the eight benchmark $LQ\bar D$ couplings of Table~\ref{tab:representative-couplings}.}
    \label{tab:slepton_decays}
\end{table}

\begin{table}[hbt!]
    \centering
    \resizebox{0.95\textwidth}{!}{
    \begin{tabular}{c c c}
    \hline\hline
    LSP  &   Coupling  &  LSP Decay   \\
    \hline\hline
    \multirow{8}{*}{$\widetilde{\ell}$} &  $\lambda_{111}'$ &\textcolor{greeen}{$2j_{\ell}$}      \\
                                   &  $\lambda_{113}'$ & \textcolor{greeen}{$1j_l + 1b$}       \\
                                   &  $\lambda_{131}'$ & \textcolor{greeen}{$1j_l + 1t$}                          \\
                                   &  $\lambda_{133}'$ & \textcolor{greeen}{$1t+1b$}                                \\
                                   &  $\lambda_{311}'$ &       $2j_l + 1l + 1\tau$~~~~~~~~$2j_l +$MET\,$^{(**)}$~~~~~~~~$2j_l + 1\tau +$MET\,$^{(**)}$ ~~~~~~~~$2j_l + 1l+$MET                 \\
                                   &  $\lambda_{313}'$ &       $1j_l +1b+ 1l + 1\tau$~~~~~~~~$1j_l+1b +$MET\,$^{(**)}$~~~~~~~~$1j_l + 1b+ 1\tau+$MET\,$^{(**)}$      ~~~~~~~~$1j_l + 1b+ 1l+$MET             \\
                                   &  $\lambda_{331}'$ & $1j_l +1t+ 1l + 1\tau$ ~~~~~~~~$1j_l+1t +$MET\,$^{(**)}$~~~~~~~~$1j_l + 1b+ 1\tau+$MET\,$^{(**)}$ ~~~~~~~~
                                  $1j_l + 1b+ 1l+$MET
                                                         \\
                                   &  $\lambda_{333}'$ &                $1t + 1b+ 1l+ 1\tau$~~~~~~~~$1t+1b+$MET\,$^{(**)}$~~~~~~~~$2b+1\tau+$MET\,$^{(**)}$~~~~~~~~$2b+1l+$MET
                                           \\
                                     
    \hline

    \hline
    \multirow{8}{*}{$\widetilde{\nu}$} &  $\lambda_{111}'$ &\textcolor{greeen}{$2j_{\ell}$}      \\
                                   &  $\lambda_{113}'$ & \textcolor{greeen}{$1j_l + 1b$}     \\
                                   &  $\lambda_{131}'$ & \textcolor{greeen}{$1j_l + 1b$}                            \\
                                   &  $\lambda_{133}'$ & \textcolor{greeen}{$2b$}                               \\
                                   &  $\lambda_{311}'$ &       $2j_l + 1l + 1\tau$\,$^{(**)}$~~~~~~~~$2j_l +$MET~~~~~~~~$2j_l + 1\tau +$MET ~~~~~~~~$2j_l + 1l+$MET\,$^{(**)}$                 \\
                                   &  $\lambda_{313}'$ &       $1j_l +1b+ 1l + 1\tau$\,$^{(**)}$~~~~~~~~$1j_l+1b +$MET ~~~~~~~~ $1j_l + 1b+ 1\tau+$MET      ~~~~~~~~$1j_l + 1b+ 1l+$MET\,$^{(**)}$             \\
                                   &  $\lambda_{331}'$ & $1j_l +1b+ 1l + 1\tau$\,$^{(**)}$~~~~~~~~$1j_l + 1b+$MET ~~~~~~~~$1j_l + 1t+ 1\tau+$MET ~~~~~~~~
                                  $1j_l + 1t+ 1l+$MET\,$^{(**)}$                         \\
                                   &  $\lambda_{333}'$ &                $2b+1l+1\tau$\,$^{(**)}$~~~~~~~~$2b+$MET~~~~~~~~$1t+1b+1\tau+$MET~~~~~~~~$1t + 1b+ 1l+ $MET\,$^{(**)}$          \\
                                     
    \hline

    \multirow{8}{*}{$\widetilde{\tau_L}$} &  $\lambda_{111}'$ &$2j_l + 1l+  1\tau$~~~~~~~~ $2j_l+$MET\,$^{(**)}$~~~~~~~~$2j_l + 1l+$MET\,$^{(**)}$~~~~~~~~$2j_l + 1\tau+$MET \\
    &$\lambda_{113}'$&$1j_l + 1b+ 1l+ 1\tau$~~~~~~~~$1j_l + 1b+$  MET\,$^{(**)}$~~~~~~~~$1j_l + 1b+ 1l+$MET\,$^{(**)}$~~~~~~~~ $1j_l + 1b+1\tau+$MET\\
    
    &$\lambda_{131}'$&$1j_l + 1t+ 1l+ 1\tau$~~~~~~~~$1j_l + 1t+$MET\,$^{(**)}$~~~~~~~~$1j_l + 1b+ 1l+$MET\,$^{(**)}$~~~~~~~~$1j_l + 1b+ 1\tau+$MET \\
    &$\lambda_{133}'$&$1t + 1b+ 1l+ 1\tau$~~~~~~~~ $ 1t+ 1b+$MET\,$^{(**)}$~~~~~~~~$ 2b+ 1l+$MET\,$^{(**)}$~~~~~~~~$ 2b+ 1\tau+$MET
    \\
    &$\lambda_{311}'$&\textcolor{greeen}{$2j_l$} \\
    &$\lambda_{313}'$&\textcolor{greeen}{$1j_l+1b$}\\
    &$\lambda_{331}'$&\textcolor{greeen}{$1j_l+1t$} \\
    &$\lambda_{333}'$&\textcolor{greeen}{$1t+1b$}\\

    \hline

    \multirow{8}{*}{$\widetilde{\nu_{\tau}}$} &  $\lambda_{111}'$ &$2j_l + 1l+  1\tau$\,$^{(**)}$~~~~~~~~ $2j_l+$MET~~~~~~~~$2j_l + 1l+$MET~~~~~~~~$2j_l + 1\tau+$MET\,$^{(**)}$ \\
    &$\lambda_{113}'$&$1j_l + 1b+ 1l+ 1\tau$\,$^{(**)}$~~~~~~~~$1j_l + 1b+$  MET~~~~~~~~$1j_l + 1b+ 1l+$MET~~~~~~~~ $1j_l + 1b+1\tau+$MET\,$^{(**)}$\\
    
    &$\lambda_{131}'$&$1j_l + 1b+ 1l+1\tau$\,$^{(**)}$~~~~~~~~$1j_l+1b+$MET~~~~~~~~$1j_l + 1t+ 1l+$MET~~~~~~~~$1j_l + 1t+ 1\tau+$MET\,$^{(**)}$ \\
    &$\lambda_{133}'$&$2b+1l+1\tau$\,$^{(**)}$~~~~~~~~$2b+$MET ~~~~~~~~$1t + 1b+ 1l+$MET~~~~~~~~ $1t + 1b+ 1\tau+$MET\,$^{(**)}$  \\
    &$\lambda_{311}'$&\textcolor{greeen}{$2j_l$} \\
    &$\lambda_{313}'$&\textcolor{greeen}{$1j_l+1b$}\\
    &$\lambda_{331}'$&\textcolor{greeen}{$1j_l+1b$} \\
    &$\lambda_{333}'$&\textcolor{greeen}{$2b$}\\
    
    \hline\hline 
    \end{tabular}
    }
    \caption{Details of the left-handed slepton LSP benchmarks with columns as in Table\,\,\ref{tab:gluino_LQD}. Here, $\widetilde{\ell} =\widetilde{e}_L/\widetilde{\mu}_L$, $\widetilde{\nu}= \widetilde{\nu}_e/\widetilde{\nu}_\mu $, $\widetilde{\ell_3} = \widetilde{\tau_L}/\widetilde{\nu_{\tau}}$.
    We show the final states from the direct decay (two-step cascade decay) in {\it \textcolor{greeen}{blue}} ({\it black}). The decay modes marked by $^{(**)}$ are mediated exclusively by charged winos ($\widetilde{W}^\pm$).} 
    \label{tab:LH_slepton_LQD}    
\end{table}

\begin{table}[hbt!]
    \centering
    \resizebox{0.7\textwidth}{!}{
    \begin{tabular}{c c c}
    \hline\hline
    LSP  &   Coupling  &  LSP Decay \\
    \hline\hline
    \multirow{8}{*}{$\widetilde{\ell}_R$} &  $\lambda_{111}'$ &$2j_l+2l$~~~~~~~~$2j_l+1l+$MET          \\
                                   &  $\lambda_{113}'$ & $1j_l+1b+2l$ ~~~~~~~~$1j_l+1b+1l+$MET          \\
                                   &  $\lambda_{131}'$ & $1j_l+1t+2l$ ~~~~~~~~$1j_l+1b+1l+$MET                       \\
                                   &  $\lambda_{133}'$ & $1t+1b+2l$ ~~~~~~~~$2b+1l+$MET                              \\
                                   &  $\lambda_{311}'$ &$2j_l+1l+1\tau$ ~~~~~~~~$2j_l+1l+$MET                \\
                                   &  $\lambda_{313}'$ &$1j_l+1b+1l+1\tau$ ~~~~~~~~$1j_l+1b+1l+$MET        \\
                                   &  $\lambda_{331}'$ & $1j_l+1t+1l+1\tau$ ~~~~~~~~$1j_l+1b+1l+$MET                      \\
                                   &  $\lambda_{333}'$ &$1t+1b+1l+1\tau$ ~~~~~~~~$2b+1l+$MET                          \\
                                     
    \hline  
    \multirow{10}{*}{$\widetilde{\tau}_R$} &  $\lambda_{111}'$ &$2j_l+1l+1\tau$~~~~~~~~$2j_l+1\tau+$MET  \\
    &$\lambda_{113}'$&$1j_l+1b+1l+1\tau$~~~~~~~~$1j_l+1b+1\tau+$MET \\
    
    &$\lambda_{131}'$&$1j_l+1t+1l+1\tau$~~~~~~~~$1j_l+1b+1\tau+$MET \\
     &$\lambda_{133}'$&$1t+1b+1l+1\tau$~~~~~~~~$2b+1\tau+$MET \\
    &$\lambda_{311}'$&$2j_l+2\tau$~~~~~~~~$2j_l+1\tau+$MET \\
    &$\lambda_{313}'$&$1j_l+1b+2\tau$~~~~~~~~$1j_l+1b+1\tau+$MET \\
    &$\lambda_{331}'$& $1j_l+1t+2\tau$~~~~~~~~$1j_l+1b+1\tau+$MET \\
    &$\lambda_{333}'$&$1t+1b+2\tau$~~~~~~~~$2b+1\tau+$MET \\
    \hline\hline 
    \end{tabular}
    }
    \caption{Details of the right-handed slepton LSP benchmarks with columns as in Table\,\,\ref{tab:gluino_LQD}. Here $\widetilde{\ell}_R =\widetilde{e}_R/\widetilde{\mu}_R$
    and we only have two-step cascade decays.}
    \label{tab:RH_slepton_LQD}    
\end{table}

Tables~\ref{tab:LH_slepton_LQD} and \ref{tab:RH_slepton_LQD} show
the final states for the different slepton and sneutrino LSPs for
each of the identified benchmark couplings. As in the case of the
left-handed squark LSPs, only one of the two possible $SU(2)_L$ 
combinations in the $LQ\bar D$ operator contributes to the direct decays of the left-handed sleptons and 
sneutrinos, resulting in a single decay mode. All cascade decays 
proceed through two-step cascades. For the left-handed sleptons 
and sneutrinos, the first step of the cascade may be mediated by 
either the bino, the neutral wino, or the charged wino, leading to additional decay modes. In contrast, the first step of the cascade for right-handed sleptons proceeds exclusively through the bino.

\subsection{Experimental Analyses at the LHC Relevant for the LQD
Couplings}
\label{sec:relevant_analyses}

In the previous section, we discussed the $LQ\bar D$ decay modes 
of the various LSPs. At colliders, however, the LSPs are 
pair-produced, and the observable signatures therefore arise from
the combination of the two LSP decays. In addition, when the LSP 
is produced through cascade decays of heavier sparticles, the 
cascades contribute extra objects to the event. In 
Ref.~\cite{Dreiner_2023}, these signatures were grouped into six 
broad classes, and the corresponding experimental analyses 
at the LHC were identified. The classification was based on
the final-state objects $L$, $j$, and $j_l$, as defined in 
Table~\ref{tab:conv}. In the present work, we refine this 
classification by introducing the previously omitted $2L+2j$ category, which arises from the direct decays of pair-produced left-handed squarks through $LQ\bar{D}$ couplings. We also generalise the $2b+2j+$MET category to $4j+$MET, where the jets may originate from light-flavour quarks, $b$-quarks, or top-quark decay products. 

Below, we present the updated classification together with the 
corresponding LHC analyses. Analyses already implemented in the 
\texttt{CheckMATE\,2} recasting framework \cite{Kim:2015wza,
Dercks:2016npn} are highlighted in \textcolor{ForestGreen}{\it 
green}.

\begin{enumerate}
    \item $4j$:
    \vspace*{-0.3cm}
    \begin{center}
    \texttt{atlas\_1710\_07171}\,\cite{ATLAS:2017jnp}, \texttt{cms\_1808\_03124}\,\cite{CMS:2018mts}, \texttt{cms\_2206\_09997}\,\cite{CMS:2022wyd},\\
    \texttt{CMS\_PAS\_EXO\_24\_039}\,\cite{CMS-PAS-EXO-24-039},
    \textcolor{ForestGreen}{\texttt{atlas\_2401\_16333}}\,\footnote{This search was implemented for our $\bar U\bar D\bar D$ study in Ref.\,\cite{Dreiner:2025kfd}.}\,\cite{ATLAS:2024kqk}\,.        
    \end{center}
    \vspace*{-0.2cm}
    Although analyses such as \texttt{cms\_2206\_09997} and \texttt{CMS\_PAS\_EXO\_24\_039} have not yet been implemented in \texttt{CheckMATE\,2}, their published experimental limits can be directly applied to our benchmark scenarios with identical signal topologies, such as $\tilde{\ell}\tilde{\ell}\to 4j_l$ for the $\lambda'_{111}$ coupling. Likewise, benchmark scenarios with $\tilde{\ell}\tilde{\ell}\to 2t+2b$ ($\lambda'_{133}$) produce high-multiplicity jet final states and are therefore expected to be probed by the ATLAS multijet search, \textcolor{ForestGreen}{\texttt{atlas\_2401\_16333}}.
    
    \item $4j+$MET:
    \vspace*{-0.3cm}
    \begin{center}
    \textcolor{ForestGreen}{\texttt{cms\_1908\_04722}}\,\cite{CMS:2019zmd}, \textcolor{ForestGreen}{\texttt{cms\_1909\_03460}}\,\cite{CMS:2019ybf},
    \textcolor{ForestGreen}{\texttt{atlas\_2004\_14060}}\,\cite{ATLAS:2020dsf},\\
    \textcolor{ForestGreen}{\texttt{atlas\_2010\_14293}}\,\cite{ATLAS:2020syg},
    \texttt{atlas\_2101\_12527}\,\cite{ATLAS:2021yij},
    \texttt{atlas\_2108\_07586}\,\cite{ATLAS:2021yqv}\,.
    \end{center}
    \vspace*{-0.2cm}
    For this class of final states, many of the relevant experimental analyses have already been implemented in \texttt{CheckMATE\,2}, making it possible to assess their sensitivity to the corresponding $LQ\bar D$ benchmark 
    scenarios. Although most of these searches were originally designed for RPC SUSY, they are also expected to be sensitive to RPV scenarios with significant missing transverse momentum
    arising from neutrinos in the final state. Some analyses, such as \texttt{atlas\_2108\_07586}\,\cite{ATLAS:2021yqv}, 
    additionally employ $W/Z/h$ tagging, which may reduce the 
    sensitivity, if the jets in the $LQ\bar D$ signal do not originate from the decay of these bosons.
    The ATLAS search \texttt{atlas\_2101\_12527}\,\cite{ATLAS:2021yij} also 
    includes an interpretation for the third-generation down-type leptoquarks, with 
    signal topologies such as $2t+2\tau$, $2b+$MET, and $1t+1b+1\tau+$MET. These topologies can be directly 
    reinterpreted for corresponding $LQ\bar D$ benchmark scenarios, for example a $\tilde{b}_R$ LSP decaying via $\lambda'_{333}$. 
    
    \item $1L+(2-6)j+$MET:
    \vspace*{-0.3cm}
    \begin{center}
    \texttt{cms\_1712\_08920}\,\cite{CMS:2017szl},
    \texttt{cms\_2012\_04178}\,\cite{CMS:2020wzx},
    \texttt{atlas\_2012\_03799}\,\cite{ATLAS:2020xzu},\\
    \textcolor{ForestGreen}{\texttt{atlas\_2101\_01629}}\,\cite{ATLAS:2021twp},
    \texttt{cms\_2102\_06976}\,\cite{CMS:2021knz}, \texttt{cms\_2107\_10892}\,\cite{CMS:2021eha},\\   
    \textcolor{ForestGreen}{\texttt{atlas\_2106\_09609}}\,\cite{ATLAS:2021fbt},
    \texttt{cms\_2208\_09700}\,\cite{CMS:2022cpe},
    \texttt{atlas\_2210\_04517}\,\cite{ATLAS:2022wcu},\\
    \texttt{cms\_2211\_08476}\,\cite{CMS:2022idi}, 
    \texttt{atlas\_2310\_08171}\,\cite{ATLAS:2023act},
    \texttt{CMS\_PAS\_SUS\_23\_014}\,\cite{CMS-PAS-SUS-23-014}\,.
    \end{center}
    \vspace*{-0.2cm}
    Among the searches in this category already implemented in \texttt{CheckMATE\,2}, \textcolor{ForestGreen}{\texttt{atlas\_2101\_01629}}\,\cite{ATLAS:2021twp} primarily targets final states arising from RPC SUSY scenarios, whereas \textcolor{ForestGreen}{\texttt{atlas\_2106\_09609}}\,\cite{ATLAS:2021fbt} is specifically designed for RPV scenarios characterised by high jet multiplicities and at least one charged lepton. Furthermore, the auxiliary information required for implementation is available in \texttt{HEPData} for several more recent searches, including \texttt{cms\_2211\_08476}\,\cite{CMS:2022idi} and \texttt{atlas\_2310\_08171}\,\cite{ATLAS:2023act}, making them good candidates for future implementation in \texttt{CheckMATE\,2}.\!\footnote{We note that both searches make use of machine learning (ML) techniques. While \texttt{HEPData} is available, additional information may be required regarding the trained models before a reliable implementation in \texttt{CheckMATE\,2} is possible.} The CMS analysis, \texttt{CMS\_PAS\_SUS\_23\_014}\,\cite{CMS-PAS-SUS-23-014}, also provides dedicated RPV interpretations for scenarios in which a bino-like neutralino or a stop LSP is produced from a sbottom or gluino NLSP, respectively, and subsequently decays via an $LQ\bar D$ coupling.
    
    \item $2L+2j$:
    \vspace*{-0.3cm}
    \begin{center}
    \texttt{CMS\_PAS\_EXO\_24\_005}\,\cite{CMS-PAS-EXO-24-005},
   \texttt{CMS\_PAS\_EXO\_24\_019}\,\cite{CMS-PAS-EXO-24-019}\,.
    \end{center}
    \vspace*{-0.2cm}
    These are leptoquark searches that yield the same signal topology and final state as some of the $LQ\bar D$ benchmark scenarios, such as $\tilde{t}_L\tilde{t}_L \to 2e+2j_l$ ($\lambda'_{131}$). The corresponding experimental limits can therefore be directly reinterpreted, even in the absence of a \texttt{CheckMATE\,2} implementation. 
    
    \item $2L+(2-6)j+$MET:
    \vspace*{-0.3cm}
    \begin{center}
    \texttt{atlas\_1710\_05544}\,\cite{ATLAS:2017jvy},
     \textcolor{ForestGreen}{\texttt{atlas\_1909\_08457}}\,\cite{ATLAS:2019fag},
     \texttt{cms\_1910\_12932}\,\cite{CMS:2019lrh},\\
     \texttt{cms\_2001\_04521}\,\cite{CMS:2020cay},
     \texttt{cms\_2001\_10086}\,\cite{CMS:2020cpy},
     \texttt{atlas\_2010\_02098}\,\cite{ATLAS:2020xov},\\
     \texttt{cms\_2012\_08600}\,\cite{CMS:2020bfa}, 
     \texttt{atlas\_2204\_13072}\,\cite{ATLAS:2022zwa},
     \texttt{cms\_2208\_09700}\,\cite{CMS:2022cpe}, \\   
     \texttt{atlas\_2305\_09322}\,\cite{ATLAS:2023lfr},
     \texttt{atlas\_2507\_00296}\,\cite{ATLAS:2025wln},
     \texttt{atlas\_2603\_16191}\,\cite{ATLAS:2026wyu},
   \texttt{CMS\_PAS\_SUS\_23\_002}\,\cite{CMS-PAS-SUS-23-002}\,.
     \end{center}
    \vspace*{-0.2cm}
    Among the searches in this class, only \textcolor{ForestGreen}{\texttt{atlas\_1909\_08457}}\,\cite{ATLAS:2019fag} has been implemented in \texttt{CheckMATE\,2}. In addition to its RPC interpretation, this analysis also presents a dedicated RPV benchmark in which a gluino decays to a top quark and a stop LSP, with the stop subsequently decaying via a $\bar U\bar D\bar D$ coupling into a bottom and a strange quark. The resulting signature consists of two leptons from the top-quark decays together with multiple jets. Furthermore, several RPC signal regions requiring significant MET may also provide sensitivity to the corresponding $LQ\bar D$ benchmark scenarios.
    
    \item $3L+4j+$MET:
    \vspace*{-0.3cm}
    \begin{center}
    \texttt{cms\_2001\_10086}\,\cite{CMS:2020cpy}, \texttt{cms\_2106\_14246}\,\cite{CMS:2021cox}, \texttt{atlas\_2307\_01094}\,\cite{ATLAS:2023afl},\\
    \texttt{atlas\_2503\_13135}\,\cite{ATLAS:2025dns}\,.
    \end{center}
    \vspace*{-0.2cm}
    In this category, \texttt{cms\_2106\_14246}\,\cite{CMS:2021cox} and \texttt{atlas\_2503\_13135}\,\cite{ATLAS:2025dns} primarily target RPC SUSY scenarios. The \texttt{cms\_2001\_10086}\,\cite{CMS:2020cpy} analysis instead considers RPV gluino LSP scenarios, including direct decays via the $\bar U\bar D\bar D$ coupling, $\tilde{g}\to tbs$, as well as five-body gluino decays yielding four jets and a charged lepton via the $LQ\bar D$ coupling. The \texttt{atlas\_2307\_01094}\,\cite{ATLAS:2023afl} analysis presents dedicated limits for scenarios in which a bino-like neutralino LSP decays via an $LQ\bar D$ coupling or a stop LSP decays via a $\bar U\bar D\bar D$ coupling, with both LSPs produced through gluino cascade decays.
    
    \item $4L+4j$:
    \vspace*{-0.3cm}
    \begin{center}
    \texttt{atlas\_2103\_11684}\,\cite{ATLAS:2021yyr}, \texttt{cms\_2106\_14246}\,\cite{CMS:2021cox}, \texttt{atlas\_2508\_19778}\,\cite{ATLAS:2025cdi}\,.
    \end{center}
    \vspace*{-0.2cm}
    Although none of these searches are currently implemented in \texttt{CheckMATE\,2}, some of the published analyses already provide dedicated RPV interpretations, such as those shown in Fig.\,11 of \texttt{atlas\_2508\_19778}~\cite{ATLAS:2025cdi}. These include scenarios with a bino-like neutralino LSP produced either in the decay of wino-like electroweakinos or from smuon decays, followed by RPV decays via the $LL\bar E$ coupling $\lambda_{12k}$ ($k=1,2$) and the $LQ\bar D$ couplings $\lambda'_{i33}$ ($i=2,3$), respectively.
\end{enumerate}

The discussion above identifies the relevant experimental analyses for the various broad classes of final states. We now quantify the resulting coverage of the $LQ\bar D$ benchmark scenarios through dedicated numerical simulations, using \texttt{CheckMATE\,2} with its available set of searches and direct reinterpretations for analyses with identical signal topologies as our benchmarks.

\section{Results}
\label{sec:results}
For our numerical analysis, we generate events using \texttt{MadGraph5\_aMC@NLO}~\cite{Alwall:2014hca} with the \texttt{RPVMSSM\_UFO}~\cite{rpvmssm} model file. The NLSP decay to LSP through gauge couplings, the RPV decays of the LSPs and the eventual showering are then handled by \texttt{Pythia\,8.2}~\cite{Sjostrand:2014zea}. The detector simulation is performed by \texttt{DELPHES\,3}~\cite{deFavereau:2013fsa} within the \texttt{CheckMATE\,2} framework. 

We now present our results in terms of 95\% confidence level (CL) mass 
exclusions. We begin with the case where LSPs are pair-produced directly at
the LHC. We then continue with the case where an NLSP cascade decays via 
gauge couplings to the LSP. The presented results are obtained using two
exclusion criteria: the default $r$-value computation in 
\texttt{CheckMATE\,2}~\cite{Dercks:2016npn}, and the $\mathrm{CL}_s$ value
from \texttt{CheckMATE\,2}'s implementation of full and simplified 
likelihoods~\cite{Lara:2025cpm}. A point on the mass parameter space is 
excluded at 95\% CL if it satisfies either of the conditions: $\mathrm{CL} 
_s<0.05$ or $r>1$.

\subsection{Direct LSP Production}
In this section, the results are based on the LSPs being pair-produced, 
which then directly decay through an $LQ\bar D$ coupling. The decay 
possibilities for various LSPs are given in 
Tables~\ref{tab:gluino_LQD}\,--\,\ref{tab:Higgsino_LQD}. We only show 
results for benchmarks with zero-step (\textit{\textcolor{greeen}{blue}} in
Tables~\ref{tab:RHsquark_LQD}, \ref{tab:LHsquark_LQD}), or through a 
one-step decay (in case of gluinos and electroweakinos, 
\textit{\textcolor{newpurple} {purple}} in Tables~\ref{tab:gluino_LQD}, 
\ref{tab:Wino_LQD}, \ref{tab:Higgsino_LQD}). As mentioned above,
in each of these scenarios all sparticles except the LSP, are considered to
be decoupled, and the LSP decays promptly. The presence of $t$- and 
$b$-jets in the final states results in higher multiplicity, which, 
depending on the targeted searches for $b$-jets or top quarks, might 
lead to a higher sensitivity. Thus, the best 
mass exclusions are often given by the $\lam'_{133}$ coupling. 
All results here are obtained by 
employing $\sqrt{s}=$13 TeV searches at the LHC,
implemented in \texttt{CheckMATE\,2}.

Due to the small production cross sections, we do not perform a numerical analysis for a direct production of bino LSP. However, we consider the possibility of pair production of
various NLSPs cascade decaying to a bino LSP in the next section, with the bino then 
decaying through an $LQ\bar D$ coupling. 

\medskip

\begin{figure}[htbp]
    \centering
    \includegraphics[width=0.8\textwidth]{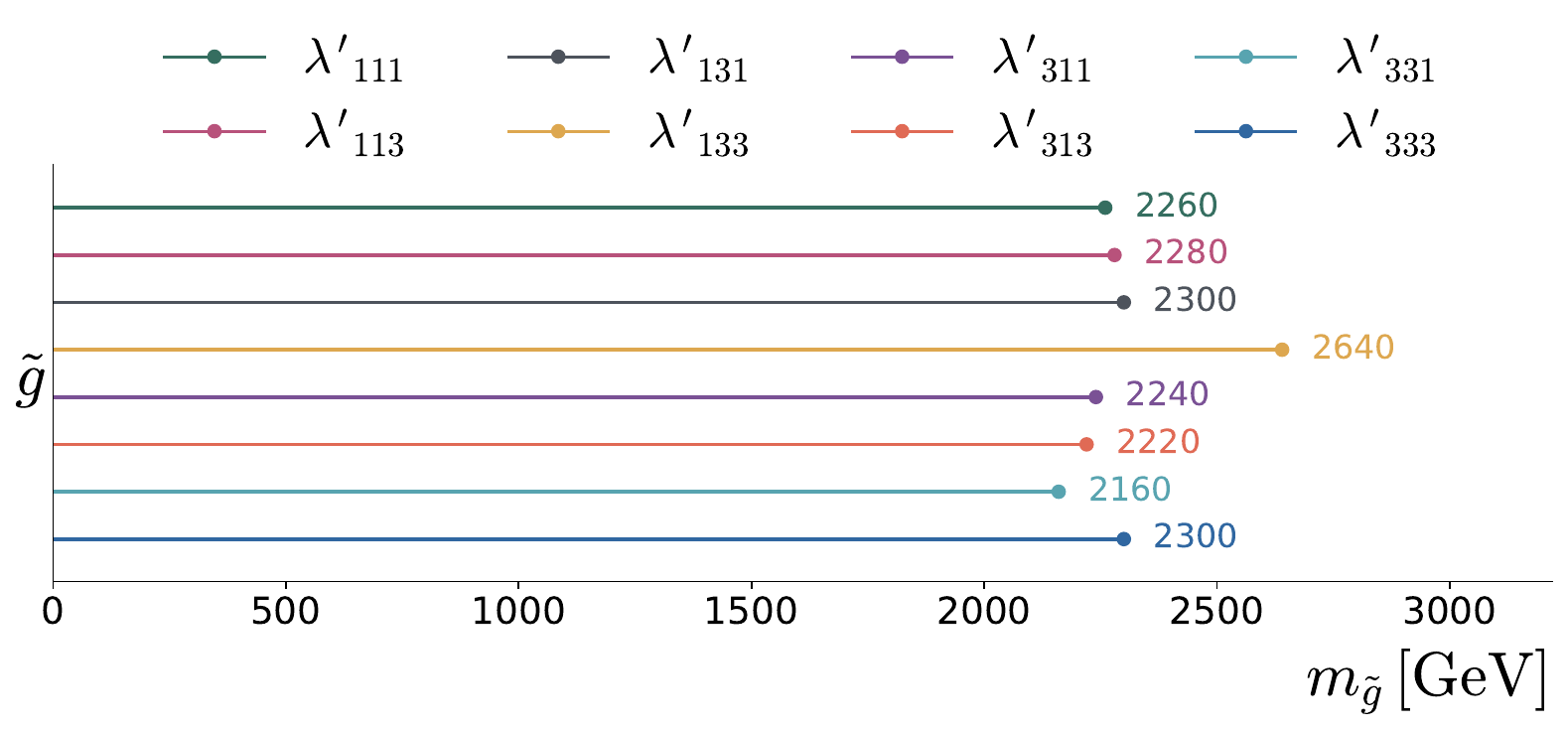}
    \caption{Search limits at 95\% CL for the direct pair production of $\tilde{g}$ LSP. The $m_{\tilde{p}}$ denotes the mass of the LSP corresponding on the y-axis. The LSP decays via $LQ\bar D$ couplings. Further details can be found in Table~\ref{tab:direct_gluino} in App.~\ref{sec:appendix-A1}. }
    \label{fig:gluino_direct}
\end{figure}

\noindent\textbf{Gluino LSP:} Due to the high pair production cross section
of gluinos at the LHC, and the implementation of likelihoods in 
\texttt{CheckMATE\,2}~\cite{Lara:2025cpm}, we obtain high mass exclusions 
for gluinos decaying through $LQ\bar D$ couplings, shown in 
Fig.~\ref{fig:gluino_direct}. Pair production cross sections were computed
using the \texttt{NNLL-fast\;2.0} computer 
code~\cite{SciPostPhysCore.7.4.072}. The results for the gluino LSP cases 
are the strongest among all the LSPs considered here, with exclusion limits
rising above 2\,TeV in all coupling scenarios. The strictest bound is for 
the $\lam'_{133}$ coupling, reaching up to $\SI{2640}{\giga\electronvolt}$.
The sensitive search for this limit is \texttt{atlas\_2211\_08028},  which searches for final states with 3 or more $b$-jets + MET.
We find that this, and many other RPC searches, are sensitive to various
$LQ\bar{D}$ benchmarks, where the MET now comes from neutrinos instead of
the stable $\tilde{\chi}_1^0$. The most sensitive search and signal region 
for each benchmark scenario are listed in 
Table~\ref{tab:direct_gluino} in App.~\ref{sec:appendix-A1}.

\medskip

\begin{figure}[htbp]
    \centering
    \includegraphics[width=0.8\textwidth]{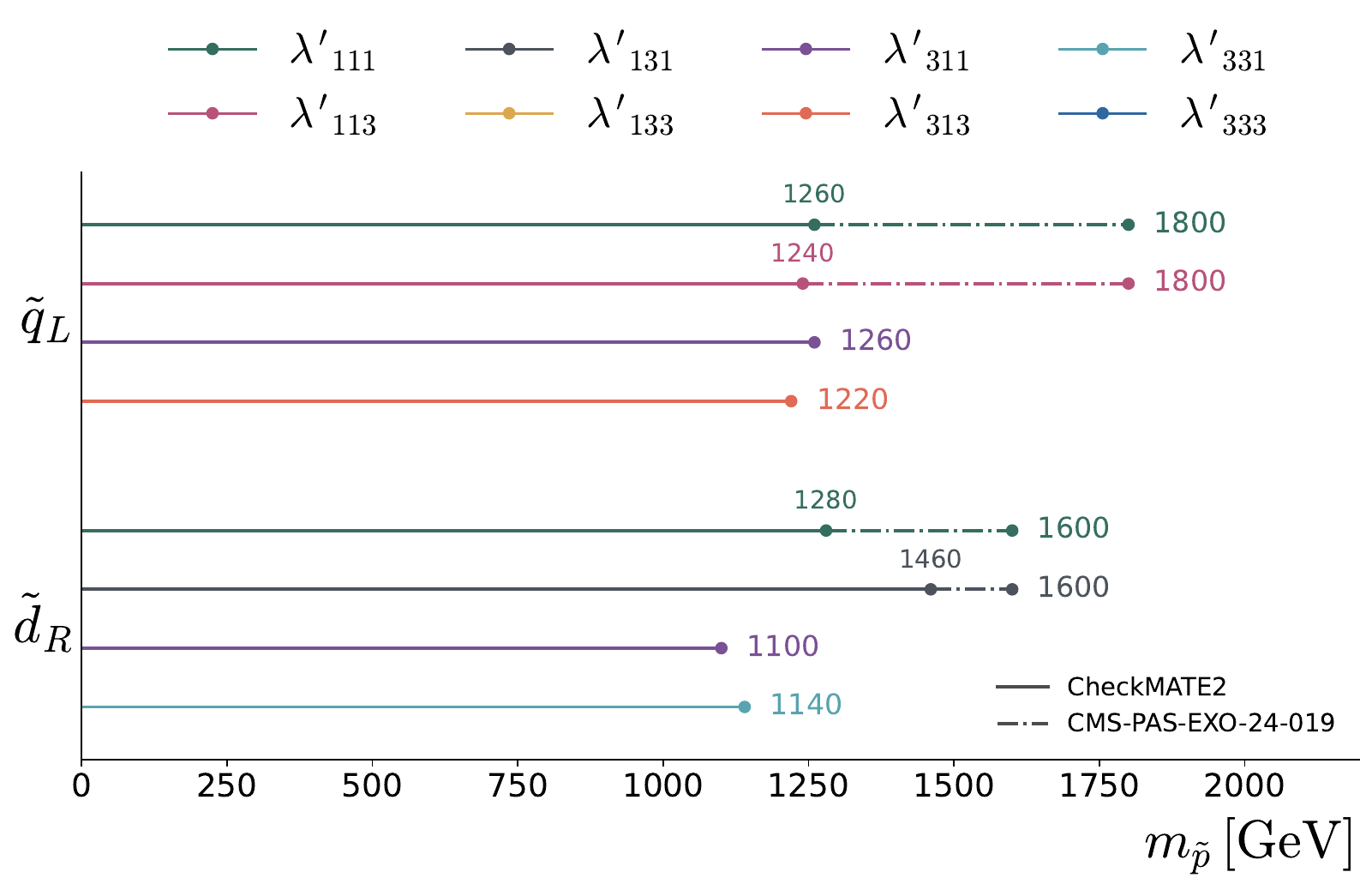}
    \includegraphics[width=0.8\textwidth]{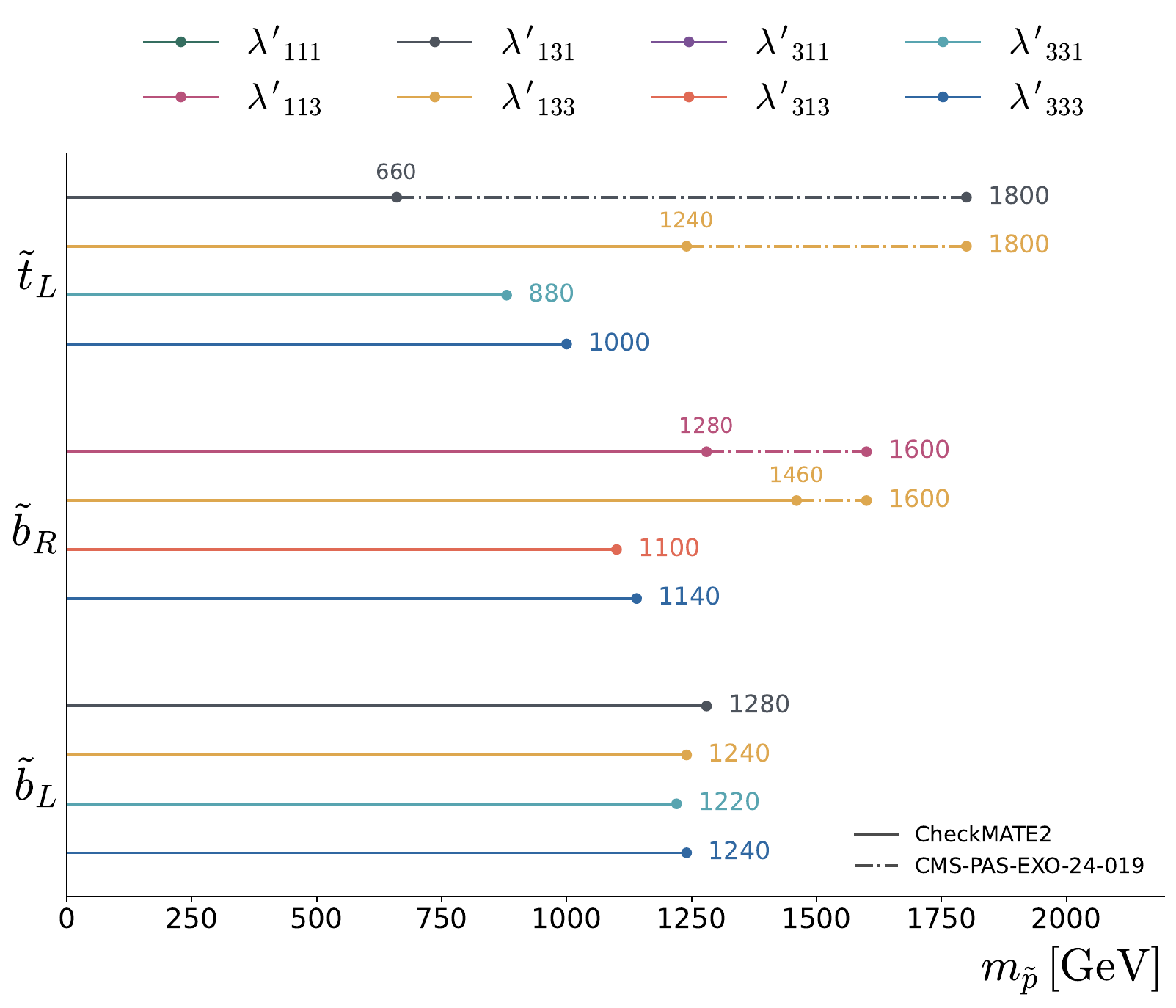}
    \caption{Search limits at 95\% CL for the direct pair
    production of a squark LSP. The $\tilde{q}_L$ denotes a
    2-fold degeneracy between ($\tilde{u}_L, \tilde{d}_L$). The
    $m_{\tilde{p}}$ denotes the mass of the LSP corresponding on the
    y-axis. The LSP decays via $LQ\bar D$ couplings. We also show
    complementary results obtained from a CMS
    search~\cite{CMS-PAS-EXO-24-019} in dot-dashed lines. The solid
    lines depict the results obtained from \texttt{CheckMATE\,2}.
    Further details can be found in Tables~\ref{tab:qL_direct},
    \ref{tab:tL_direct}, \ref{tab:bR_direct}, and 
    \ref{tab:bL_direct} in App.~\ref{sec:appendix-A1}.}
    \label{fig:squark_direct}
\end{figure}

\noindent
\textbf{Squark LSPs:} We only show results for a selected 
number of squark LSPs, where direct decays are possible. These 
are shown in \textit{\textcolor{greeen}{blue}} in 
Tables~\ref{tab:RHsquark_LQD}  and \ref{tab:LHsquark_LQD}. The 
production cross sections for squarks are computed assuming a 10-fold degeneracy using the \texttt{NNLL-fast\;2.0}
code~\cite{SciPostPhysCore.7.4.072}. The results are shown in 
Fig.~\ref{fig:squark_direct}. Here, the $\tilde{q}_L$ benchmark 
assumes a 2-fold 
degeneracy between ($\tilde{u}_L,\,\tilde{d}_L$). In every other 
benchmark, the given squark is assumed to be the only 
non-degenerate LSP. In cases where two decay modes for the LSP 
are possible with the same $LQ\bar D$ coupling, we  consider a 50\% 
branching ratio to each decay mode, since they are mediated by the same $LQ\bar D$ coupling. The different LSPs show 
varying sensitivity. This can be attributed to the 
presence/absence of MET, $t-$ or $b-$quarks in the final state. 
The results for $\tilde{d}_R$ are similar to those of 
$\tilde{b}_R$, where the relevant couplings change to match the 
final states.

The results obtained for these models from \texttt{CheckMATE\,2} are shown 
as solid lines (more detailed information is given in 
Tables~\ref{tab:qL_direct},\,\ref{tab:tL_direct},\,\ref{tab:bR_direct}, and
\ref{tab:bL_direct}), while the dot-dashed lines show results obtained from
a CMS search targeting leptoquarks resulting in a $2l+2j_l$ final 
state~\cite{CMS-PAS-EXO-24-019}. This search also includes the case where 
the final state can include $b-$quarks, i.e. a $2l+2b$ final state. Thus, 
the result from this CMS search can be directly translated to the 
benchmarks $\tilde{q}_L (\lam'_{111},\,\lam'_{113})$, $\tilde{d}_R (\lam'_ 
{111},\,\lam'_{131})$, $\tilde{t}_L (\lam'_{131},\,\lam'_{133})$, and 
$\tilde{b}_R (\lam'_{113},\,\lam'_{133})$. For benchmarks where the 
$2l+2j_l$ final state is one of the three possible final-states, we scale 
the pair-production cross section by its corresponding probability, i.e. 
25\%.  We compare this with the experimental upper limit on the 
corresponding cross sections from the CMS  
search~\cite{CMS-PAS-EXO-24-019}. 
The mass exclusions obtained from the CMS search~\cite{CMS-PAS-EXO-24-019} 
outperform every corresponding result obtained from 
\texttt{CheckMATE\,2}, with the strictest bounds reaching up to $\SI{1800} 
{\giga\electronvolt}$ for the $\tilde{q}_L$ ($\lam'_{111},\, \lam'_{113})$ 
and $\tilde{t}_L (\lam'_{131},\,\lam'_{133})$ cases. The strictest limit 
from \texttt{CheckMATE\,2} is for a $\tilde{d}_R $/$\tilde{b}_R$ LSP 
for the coupling $\lam'_{131}/\lam'_{133}$ respectively, reaching 
$\SI{1460}{\giga\electronvolt}$.

\medskip

\noindent
\textbf{Electroweakino LSP:} Due to small pair production 
cross sections of $\widetilde{B}$, we do not perform the numerical study for direct production of bino LSP, we only study its production via cascade decay of another sparticle. However, using 
\texttt{CheckMATE\,2} we obtain mass exclusions for winos and 
higgsinos. For wino-like LSPs, we assume that $\tilde{\chi}^0 
_1$ and $\tilde{\chi}^{\pm}_1$ are mass degenerate, and are
produced at the LHC as $pp\to\tilde{\chi}^0_1\tilde{\chi}^{\pm} 
_1$ and $pp\to\tilde{\chi}^{+}_1\tilde{\chi}^{-}_1$. For higgsino-like
LSPs, the mass degeneracy is assumed between $\tilde{\chi}^0
_1$, $\tilde{\chi}^0_2$ and $\tilde{\chi}^{\pm}$ and are produced
as $pp\to \tilde{\chi}^0_1 \tilde{\chi}^{\pm}_1,\, \tilde{\chi}^0_2
\tilde{\chi}^{\pm}_1 ,\, \tilde{\chi}^0_1 \tilde{\chi}^0_2 ,\,  \tilde{\chi}^{+}_1 \tilde{\chi}^{-}_1\,$. The production 
cross sections are obtained employing  the 
\texttt{Resummino} code~\cite{Fuks:2012qx, Fuks:2013vua} with the CTEQ6.6 and MSTW2008nlo90cl PDFs~\cite{Nadolsky:2008zw,Martin:2009iq}.

\begin{figure}[ht!]
    \centering
    \includegraphics[width=0.8\textwidth]{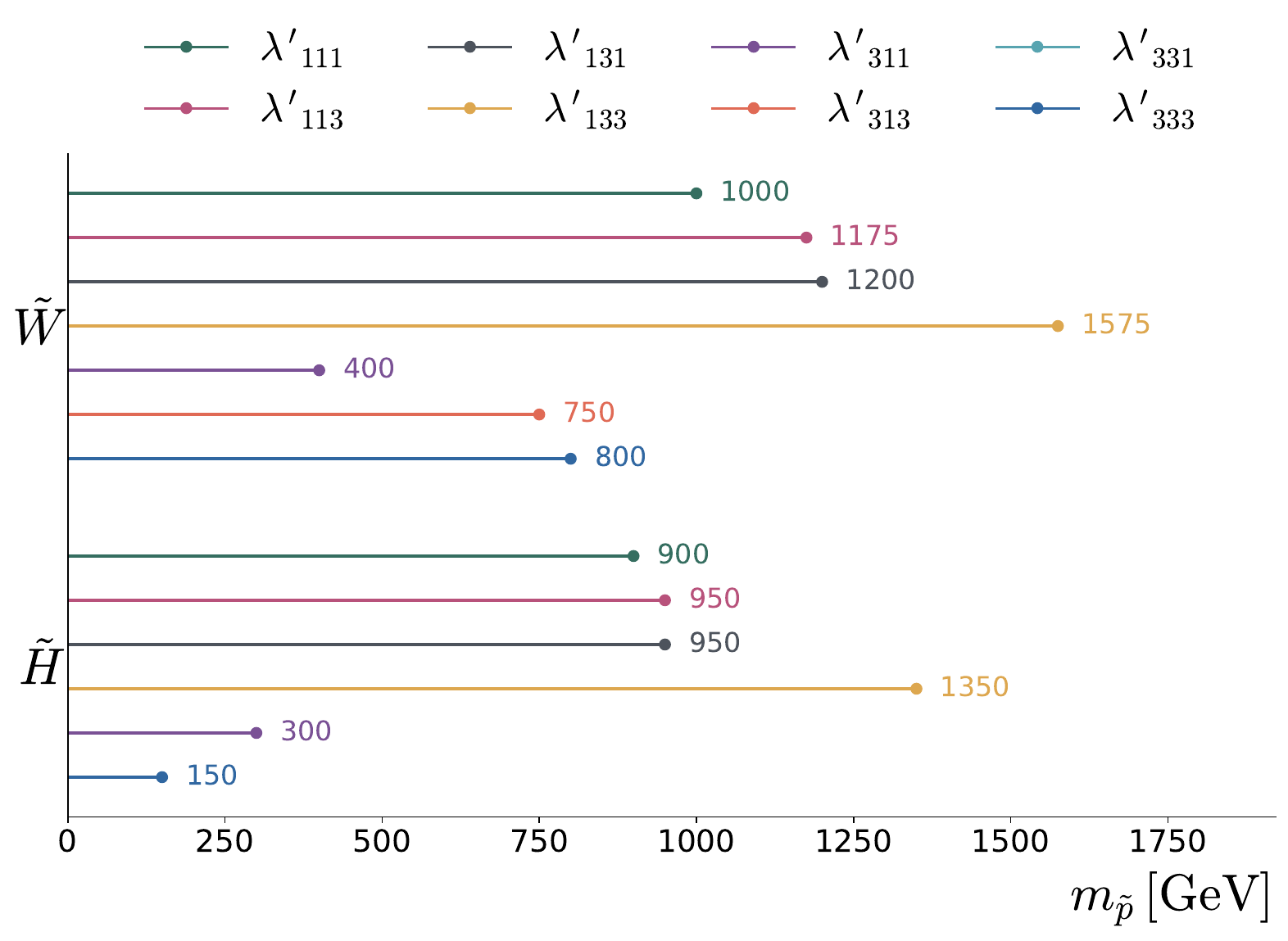}
    \caption{Search limits at 95\% CL for the direct pair production of electroweakino LSPs. The $m_{\tilde{p}}$ denotes the mass of the LSP corresponding on the y-axis. The LSP decays via $LQ\bar D$ couplings. Further details can be found in Tables~\ref{tab:W_direct} and \ref{tab:H_direct} in App.~\ref{sec:appendix-A1}. }
    \label{fig:ewino_direct}
\end{figure}

The results obtained for these benchmarks are shown in 
Fig.~\ref{fig:ewino_direct}, while more detailed information is provided in
Tables~\ref{tab:W_direct} and \ref{tab:H_direct}. For some coupling 
benchmarks, \texttt{CheckMATE\,2} provides no exclusion from any 
implemented searches. These are the couplings $\lam'_{331}$ for wino LSPs 
and $\lam'_{313}, \lam'_{331}$ for higgsino LSPs. As in previous benchmark 
scenarios, $\lam'_{133}$ leads to the largest mass exclusion of all couplings for both winos and 
higgsinos, reaching a maximum of $\SI{1575}{\giga 
\electronvolt}$ for $\widetilde{W}$ LSP. The sensitive search for this 
limit is \texttt{atlas\_2211\_08028},  which looks for final states with 3 
or more $b$~jets + MET. Weaker limits are obtained for couplings $\lam'
_{3ij}$, which can be attributed to the weak reconstruction of $\tau$ in 
the final state.

\medskip

\noindent
\textbf{Slepton LSP:} For slepton LSPs, we find no mass exclusions for any benchmark couplings with \texttt{CheckMATE\,2}. For the final states with 4 jets involving $t$-jets, we also check the sensitivity of the ATLAS multijet search~\cite{ATLAS:2024kqk} implementation in \texttt{CheckMATE\,2}, which also does not yield any sensitivity. This can be a combined effect of the high jet energy thresholds required for the ATLAS multijet search, and the low pair-production cross sections for sleptons. We attempted to compare with the result from the CMS dijet searches~\cite{CMS-PAS-EXO-24-039,CMS:2022wyd}, since some of the slepton LSP scenarios have similar final states with $4j_l$ or $2j_l+2b$. However, due to the low pair-production cross sections of the sleptons, we obtained no exclusion. Slepton LSPs produced through an NLSP could have sensitivity from existing LHC searches due to higher multiplicity of final states -- we leave this scenario for a future work.

\subsection{Bino LSP Production through the Cascade Decay of an NLSP}
\label{sec:cascadedecay}

An LSP can also be produced through an NLSP, where the NLSP
is pair-produced at the LHC. The cascade decay of a heavier NLSP leads to the production of the lighter LSP, 
which can then further decay via an $LQ\bar D$ 
coupling. Due to the higher multiplicity of final
states (with extra leptons/jets, depending on the NLSP), cascade 
decay benchmarks can have a higher sensitivity to the ATLAS/CMS searches than in the 
direct LSP production case. Here, we focus on the bino LSP produced 
through an NLSP, since we have not studied the direct production of the bino LSP. For the squark and slepton NLSPs, we focus specifically on
the couplings where the NLSP mediates the bino LSP decay through the $LQ\bar D$
coupling. We present results as 95\% CL exclusion contours over the
NLSP-LSP mass plane.  


\subsubsection*{Gluino NLSP}
\label{subsec:gluinotobino}
\begin{figure}[htbp!]
    \centering
    \begin{minipage}{0.49\textwidth}
    \includegraphics[width=\textwidth]{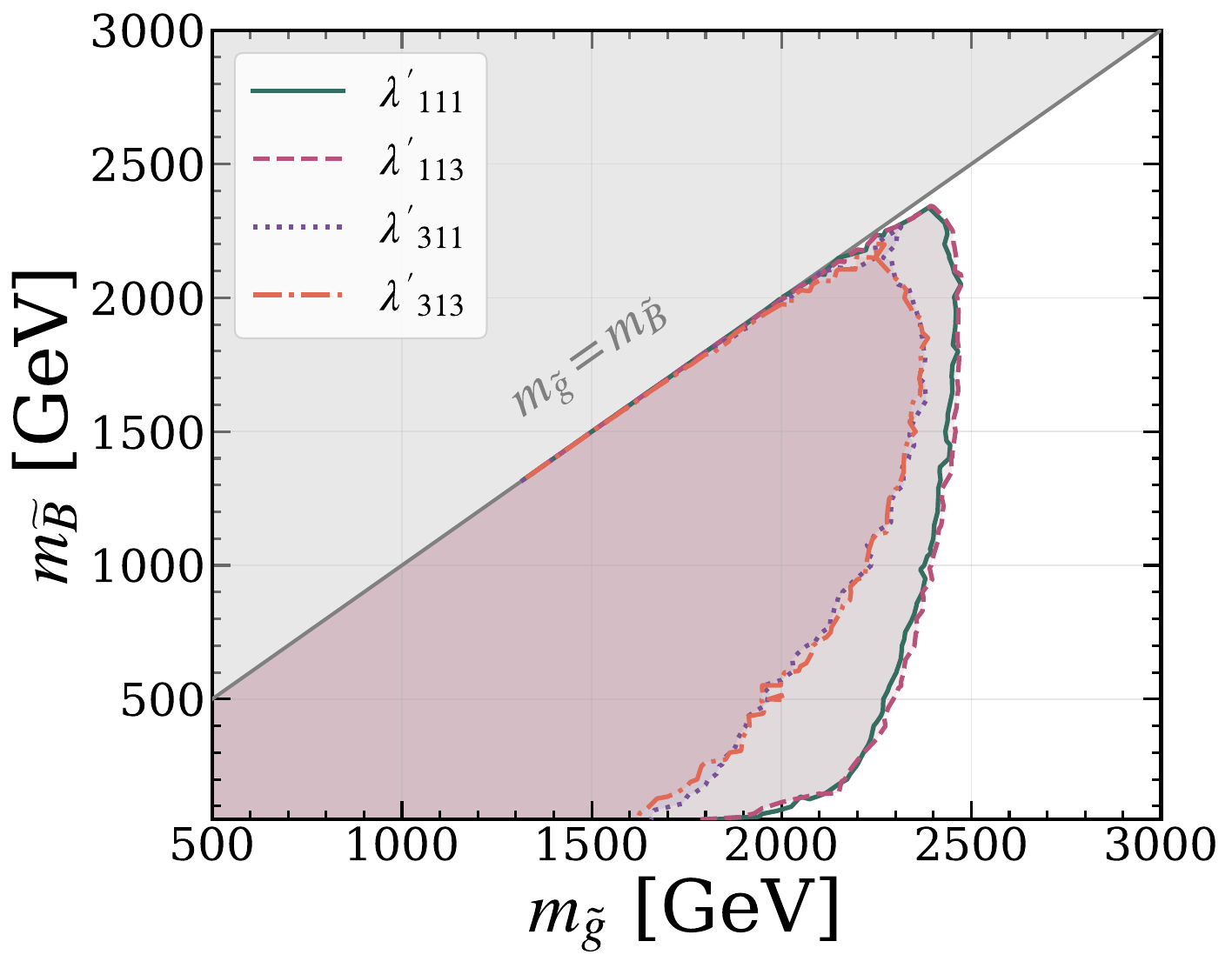}
    \end{minipage}
    \begin{minipage}{0.49\textwidth}
    \centering
    \includegraphics[width=\textwidth]{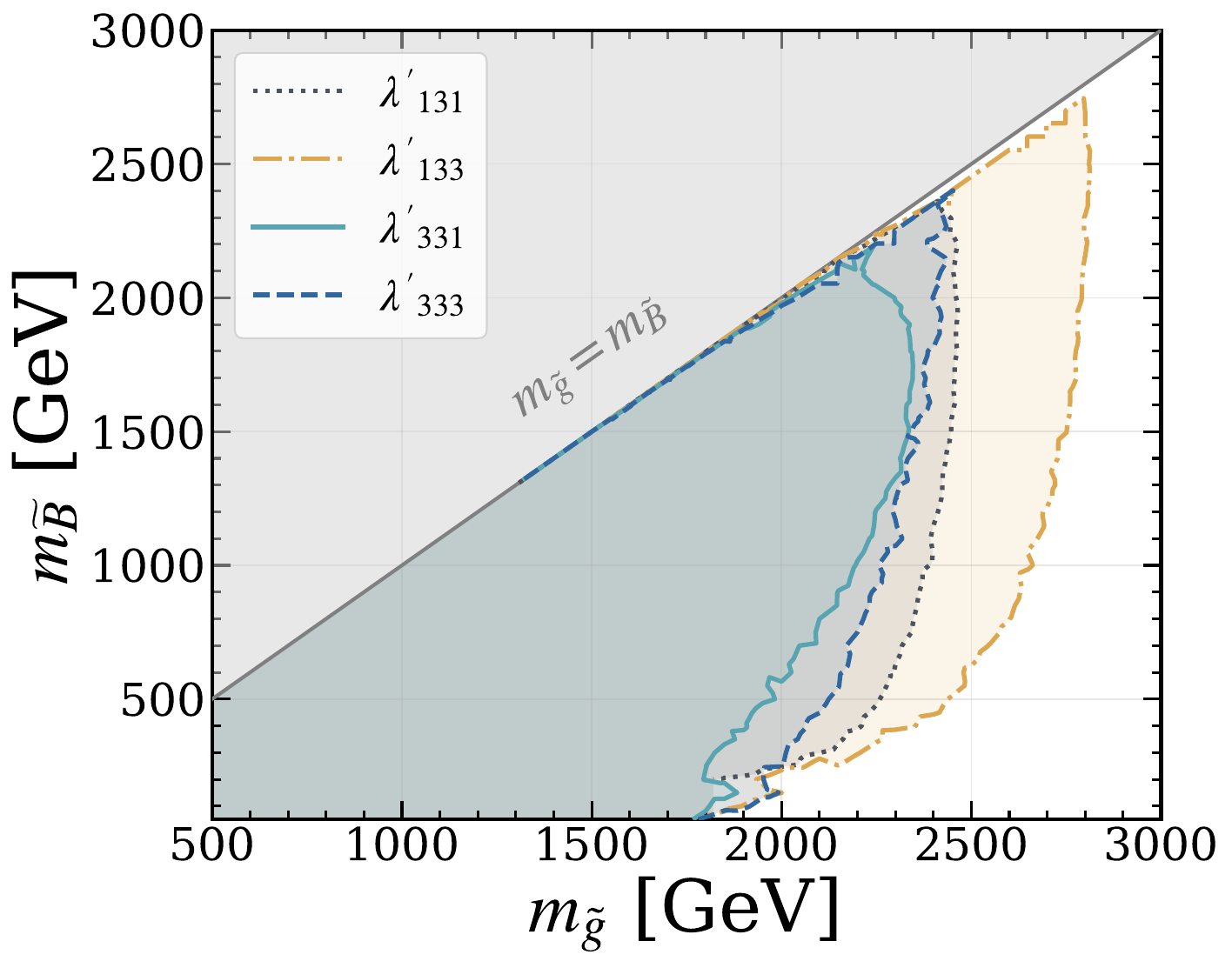}
    \end{minipage}
    \caption{Exclusion regions for the case of an NLSP gluino cascade decaying to an LSP bino as a function of the gluino (x-axis) and bino (y-axis) masses, respectively. 
    The \textit{left} panel shows results for $LQ\bar D$ couplings that yield
    light jets + leptons, while the \textit{right} panel for couplings yielding final states with top jets. The  gray region in both panels depicts the 
    kinematically inaccessible region. A list of the sensitive searches 
    can be found in Table~\ref{tab:gluinotobino}
    of App.~\ref{sec:appendix-A2}.}
    \label{fig:gluinotobino}
\end{figure}

\begin{figure}[htbp!]
    \centering
    \begin{minipage}{0.49\textwidth}
    \includegraphics[width=\textwidth]{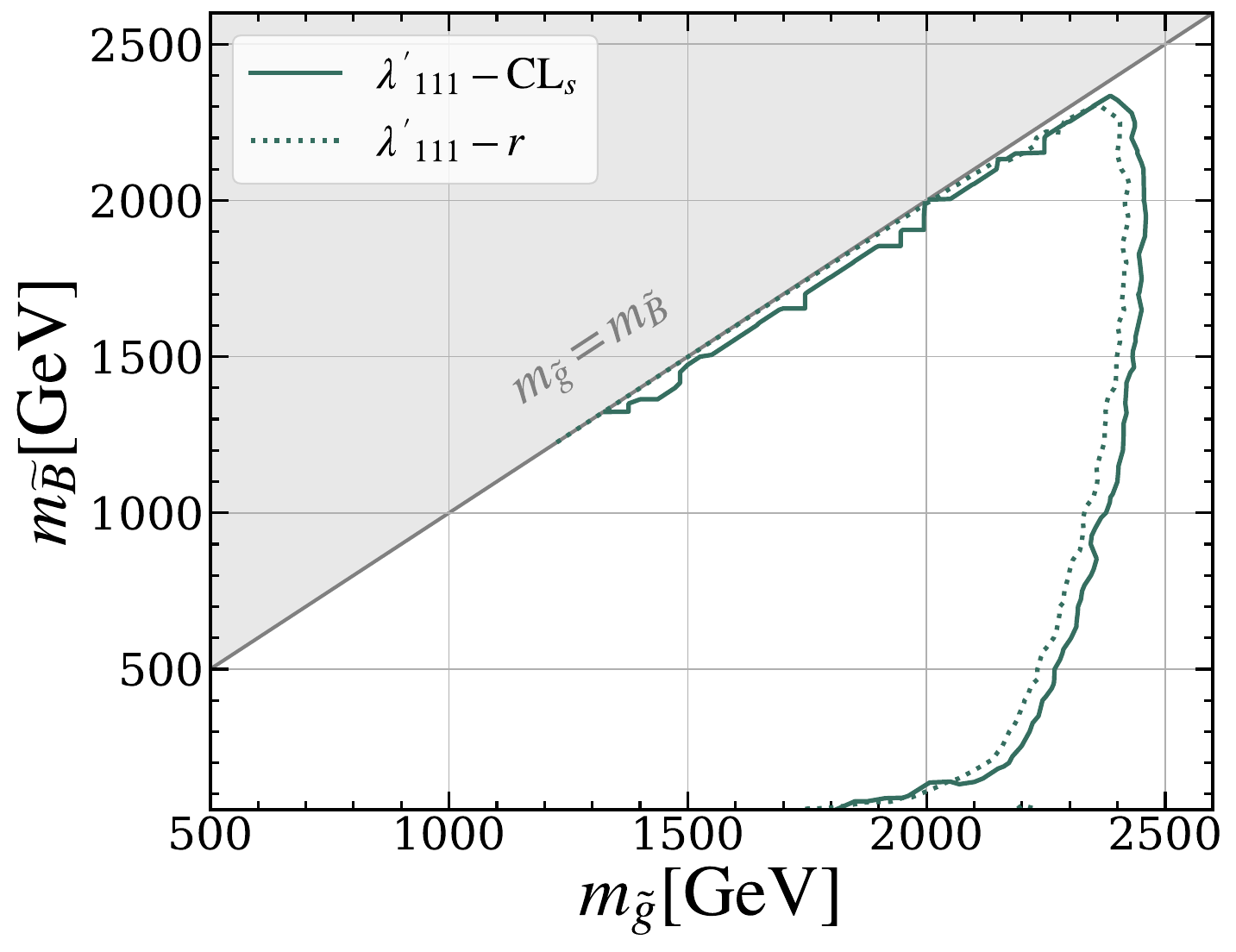}
    \end{minipage}
    \begin{minipage}{0.49\textwidth}
    \centering
    \includegraphics[width=\textwidth]{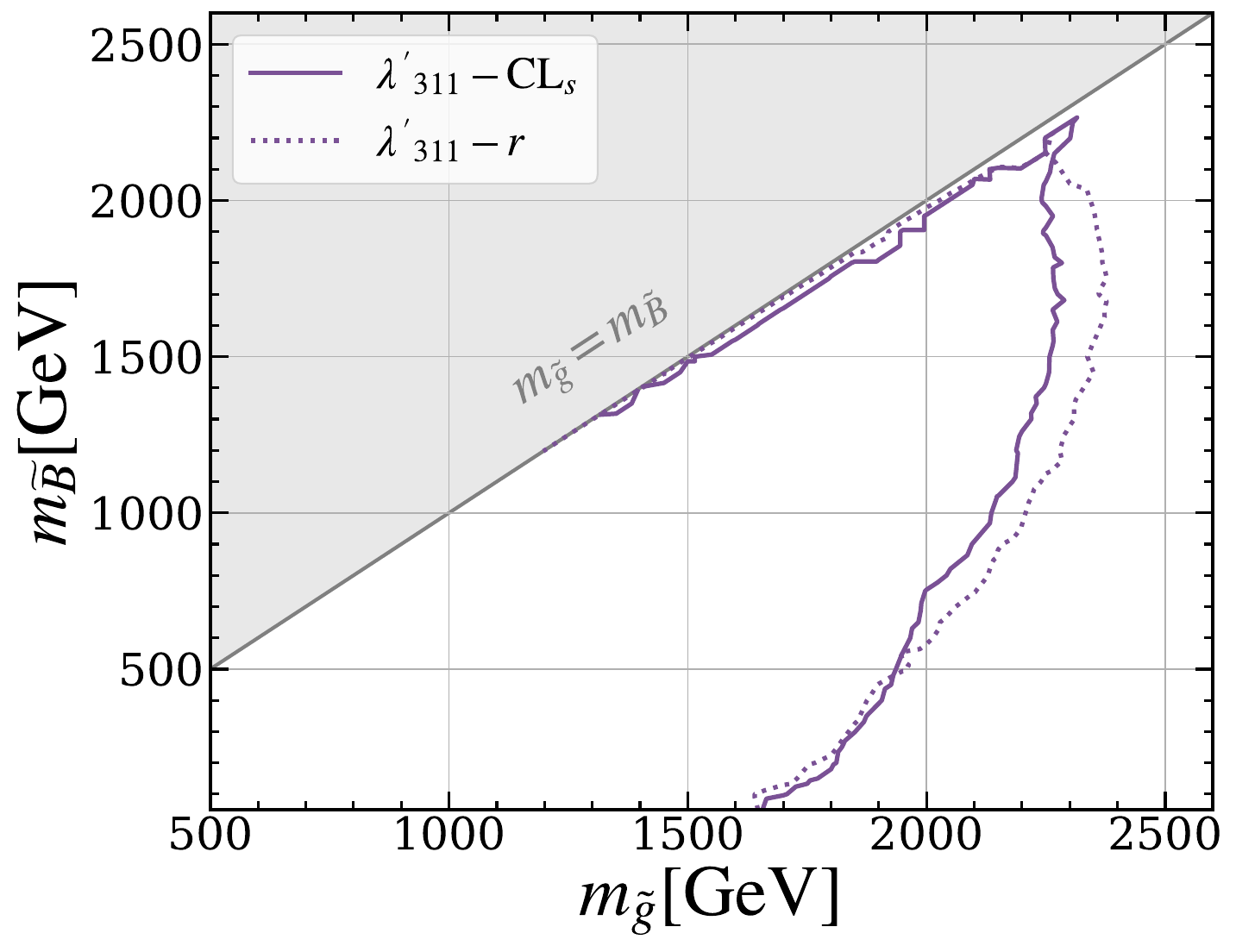}
    \end{minipage}
    \caption{Exclusion regions for the case of an NLSP gluino 
    cascade decaying to an LSP bino as a function of the gluino 
    and bino masses. We consider the couplings $\lam'_{111}$ on 
    the \textit{left} and $\lam'_{311}$ on the \textit{right}. We furthermore show
    the difference between two types of results from  
    \texttt{CheckMATE\,2}, $\mathrm{CL}_s$ vs. $r$. The gray region
    in both panels depicts the kinematically inaccessible region. The
    solid line shows the exclusion obtained from the likelihood 
    implementation of \texttt{CheckMATE\,2}, which corresponds to
    $\mathrm{CL}_s < 0.05$. The dotted line shows the mass exclusion 
    obtained from the default version of
    \texttt{CheckMATE\,2}. This corresponds to $r>1$. Their 
    combination leads to a larger exclusion in the mass parameter space, as shown in Fig.~\ref{fig:gluinotobino}.}
    \label{fig:gluinotobino_difference}
\end{figure}

Since gluino pair production is expected to have a high cross section at
the LHC, we consider all eight $LQ\bar D$ couplings for the $\tilde{g}\to 
\widetilde{B}+X$ cascade benchmark. The scan is performed over the 
gluino and bino masses with the condition $m_{\tilde{g}}\geq m_ 
{\widetilde{B}}$. The results are presented in 
Fig.~\ref{fig:gluinotobino}. The left panel shows results for final 
states with light jets+leptons, while the right panel shows
results for final states with light jets+leptons along with top jets. Fig.~\ref{fig:gluinotobino_difference} shows the difference in mass 
exclusion one obtains if only either of the $r-$value or CL$_s$ value approaches in \texttt{CheckMATE\,2} is used. We 
perform a Delaunay triangulation~\cite{Delaunay:1934}, as
implemented in the \texttt{matplotlib.tri} module~\cite{Hunter:2007}, in order to interpolate the results 
obtained from both approaches, shown in Fig.~\ref{fig:gluinotobino}, 
and all our results henceforth. A combination of both approaches yields
a larger exclusion in the $m_{\tilde{g}}- m_{\widetilde{B}}$ mass plane.

\begin{figure}[htbp!]
    \centering
    \includegraphics[width=0.6\textwidth]{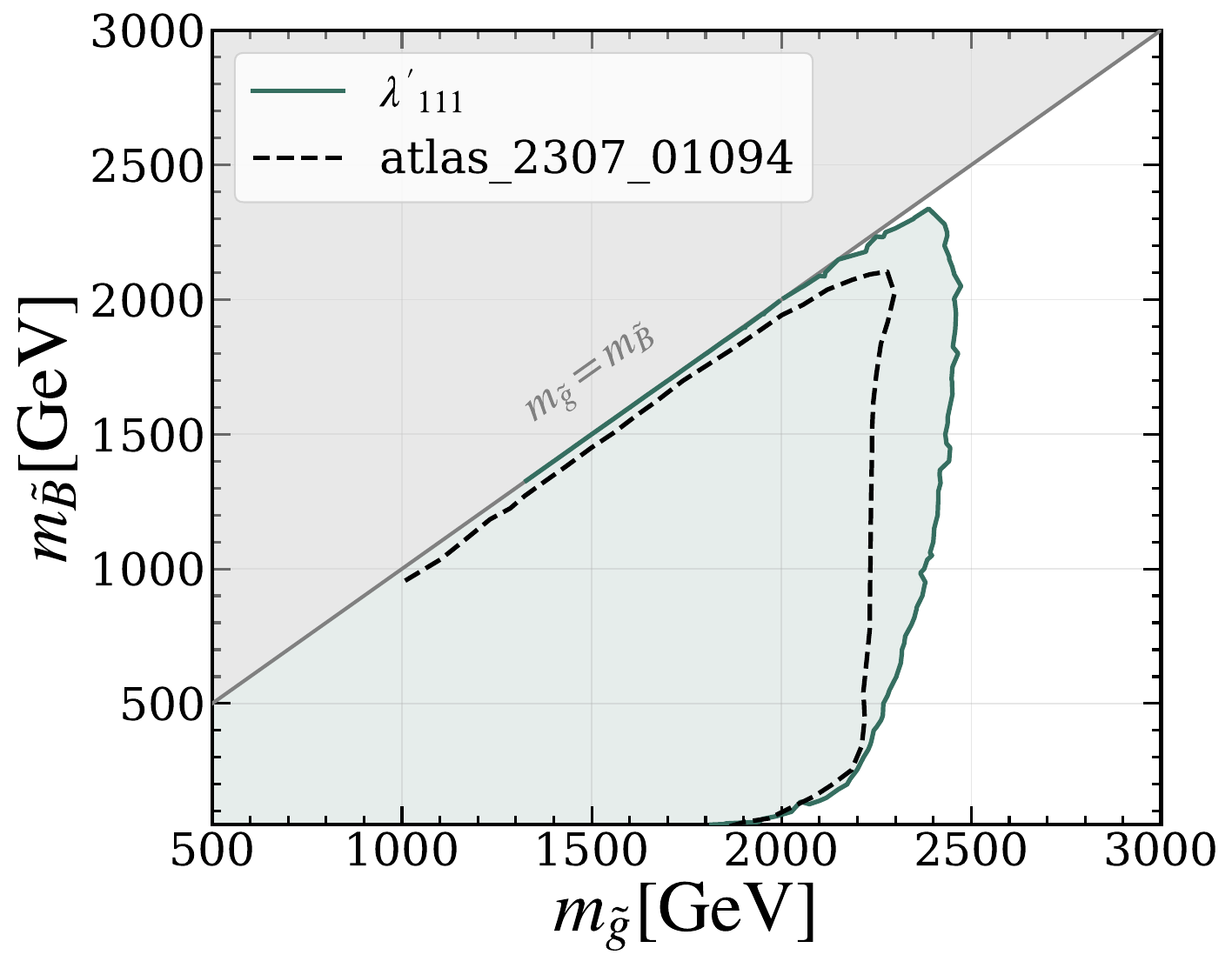}
    \caption{Comparison of the $\tilde{g}$ to $\widetilde{B}$  
    cascade results for $\lam'_{111}$ from \texttt{CheckMATE\,2} 
    (solid line) with Fig.~7(e) of 
    \texttt{atlas\_2307\_01094}~\cite{ATLAS:2023afl} (dashed).}
    \label{fig:gluinotobino_atlascomparison}
\end{figure}

As in the direct production benchmarks, the largest exclusion in 
the mass plane is achieved in the $\lam'_{133}$ case, reaching up to $\SI{2.8}{\tera\electronvolt}$ in $m_{\tilde{g}}$ for a nearly equal bino mass. The presence of $\tau$ 
leptons in the final state tends to reduce sensitivity due to their
lower reconstruction and identification efficiency. This results in 
smaller mass exclusion contours for  $\lam'_{3jk}$ compared to other 
couplings. We list the searches that contribute sensitivity to this 
benchmark across all couplings, in  
Table~\ref{tab:gluinotobino} in App.~\ref{sec:appendix-A2}. They are ordered by relevance, where they appear in decreasing order of the number of mass points they exclude. This is inclusive of both approaches of \texttt{CheckMATE\,2}, based on $r$- and CL$_s$-values. We furthermore compare our 
results to a similar cascade decay studied by an ATLAS search 
involving the $\lam'_{111}$ 
coupling~\cite{ATLAS:2023afl}, but is not currently implemented in the \texttt{CheckMATE\,2} framework. The comparison is shown in 
Fig.~\ref{fig:gluinotobino_atlascomparison}. Our results from the searches already implemented in
\texttt{CheckMATE\,2} yield a stronger exclusion across the mass plane 
than the ATLAS result, but are qualitatively in agreement.

\vspace{0.5cm}

\subsubsection*{Light Flavour Squark NLSP}

\begin{figure}[htbp!]
    \centering
    \begin{minipage}{0.49\textwidth}
    \includegraphics[width=\textwidth]{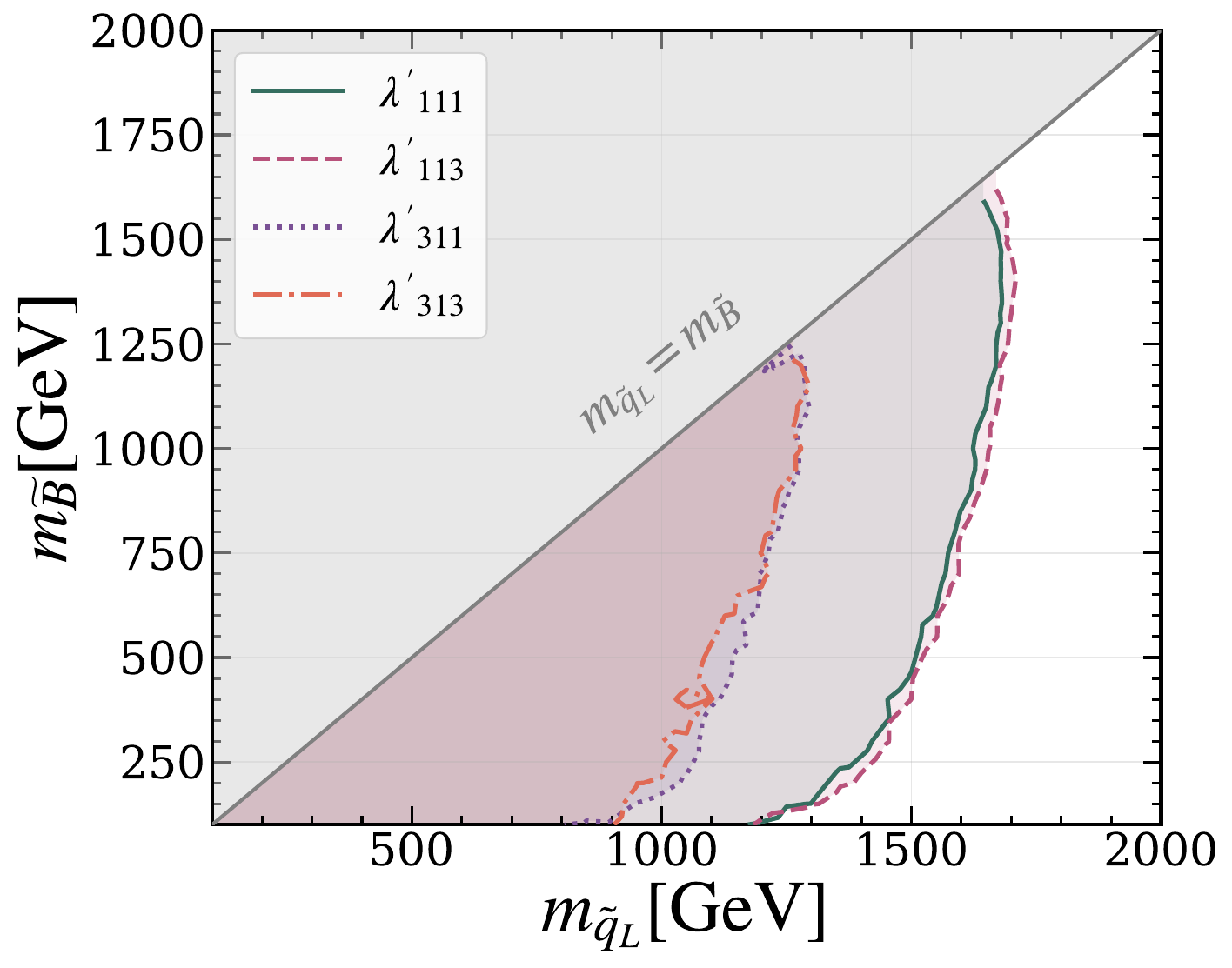}
    \end{minipage}
    \begin{minipage}{0.49\textwidth}
    \centering
    \includegraphics[width=\textwidth]{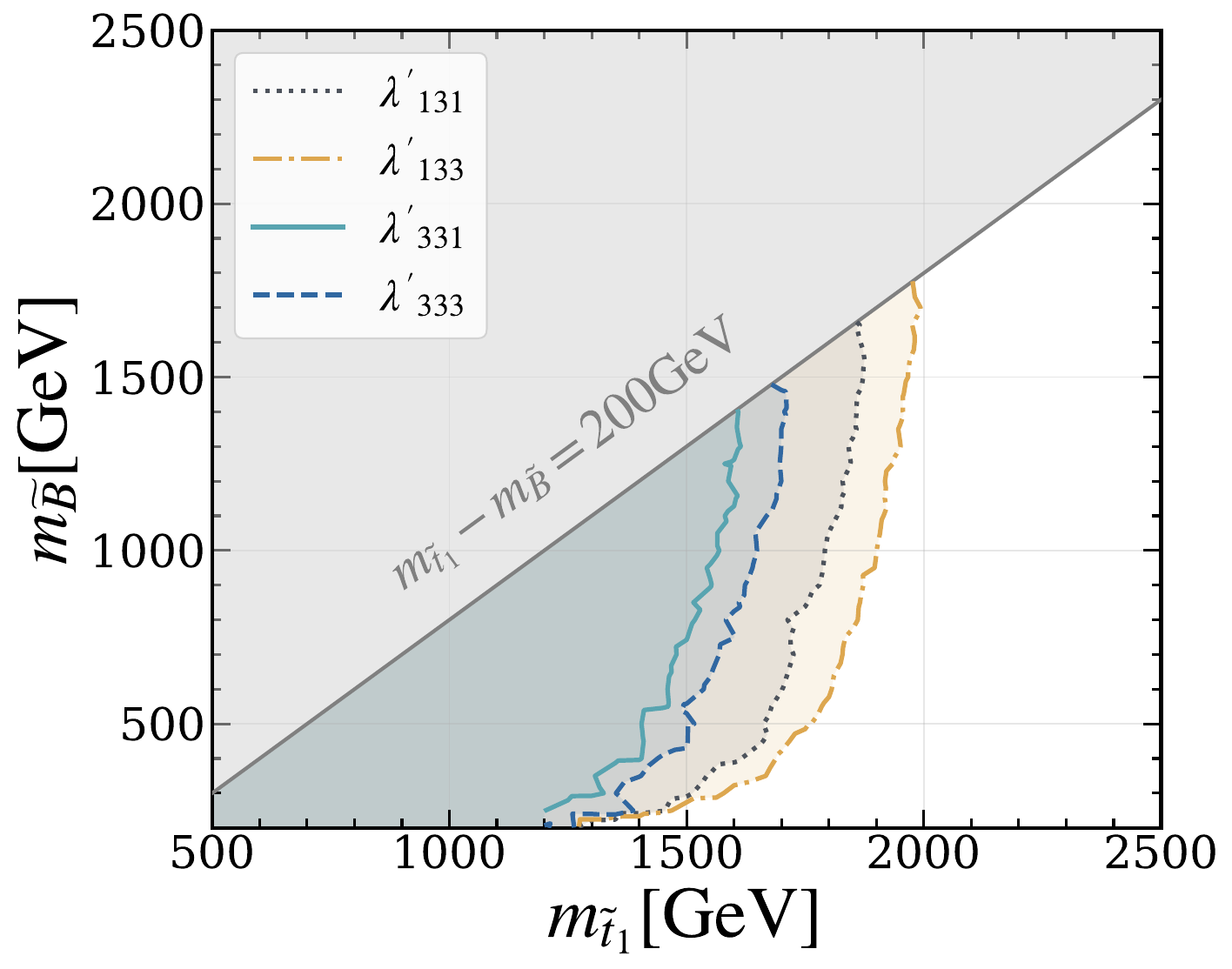}
    \end{minipage}
    \caption{\textit{Left}: Results for the cascade decay of a light
    squark, $\tilde{q}_L$, to a bino, $\widetilde{B}$, where $\tilde{q} 
    _L = (\tilde{u}_L,\tilde{d}_L)$ is the NLSP and 
    $\widetilde{B}$ the LSP. \textit{Right}: Results for the 
    cascade decay of a NLSP stop, $\tilde{t}_1$, to an LSP
    $\widetilde{B}$. The gray region depicts the kinematically 
    inaccessible region. A list of sensitive searches can be found in 
    Table~\ref{tab:squarktobino} of App.~\ref{sec:appendix-A2}.}
    \label{fig:squarktobino}
\end{figure}

In this benchmark where a light-flavour squark NLSP, $\tilde{q}_L$,
cascade decays to a bino $\widetilde{B}$, we consider the couplings 
$\lam'_{111}, \lam'_{113},\lam'_{311}$ and $\lam'_{313}$. The gluinos are
considered to be decoupled, and hence the pair production of squarks is 
flavour-independent. We consider a two-fold mass degeneracy between 
($\tilde{u}_L, \tilde{d}_L$), and use the 10-fold production cross 
sections from the \texttt{NNLL-fast~2.0} computer code scaled down 
by a factor of five, for a two-fold degeneracy. We scan over sparticle 
masses with the condition $m_{\tilde{q}_L}\geq m_{\widetilde{B}}$. 
The left panel of Fig.~\ref{fig:squarktobino} shows the results as an 
exclusion contour in the ($m_{\tilde{q}_L}, m_ {\widetilde{B}}$) mass plane. As in the previous benchmark, the result is obtained through interpolation between the results of \texttt{CheckMATE\,2}. Similar to the previous benchmark, couplings $\lam'_{31j}$ yielding $\tau$ leptons in the final state have a smaller exclusion region compared to the case of $e,\,\mu$.

\medskip

\subsubsection*{Stop NLSP}

In this cascade scenario, we consider the couplings $\lam'_{131},\, \lam'
_{133},\,\lam'_{331}$ and $\lam'_{333}$ for an NLSP stop $\tilde{t}_1$ 
decaying into the bino $\widetilde 
{B}$ LSP, such that $\tilde{t}_1\to t+\widetilde 
{B}$. Thus we employ $m_{\tilde{t}_1}-m_{\widetilde{B}}\geq m_t$ throughout. Here, we only consider the 
lighter stop, $\tilde{t}_1$, in the spectrum, and assume $\tilde{t}
_2$ to be decoupled. We also assume minimal mixing between the two, such 
that $\tilde{t}_1$ is dominantly left-handed~\cite{Martin:1997ns}. This also assists the $LQ\bar D$ decay of the bino LSP via the left-handed stop NLSP as the intermediate sparticle. The 
pair production cross sections were obtained using
\texttt{NNLL-fast~2.0}. Fig.~\ref{fig:squarktobino} (right) shows the 
results obtained from \texttt{CheckMATE\,2} with similar conditions for 
the exclusion contour as for the previous benchmark, now for 
the mass plane ($m_{\tilde{t}_1},m_{\widetilde{B}}$). We find a larger exclusion region in this mass plane for this set of 
couplings as compared to the ${\tilde{q}_L}$ to ${\widetilde{B}}$ 
cascade. This can be attributed to the extra $t$ jets in the final state.
$\lam'_{133}$, as before, has the largest exclusion. We list the searches
that contribute sensitivity to both benchmarks across all couplings, 
ordered by relevance, in Table~\ref{tab:squarktobino} in 
App.~\ref{sec:appendix}.

\subsubsection*{Slepton NLSP}
\label{sec:sleptontobino}
We now consider the benchmark where with a light-flavoured slepton
NLSP, $\tilde{e}_L$, cascade decays to a bino, $\widetilde{B}$. We 
perform the scan with the condition $m_{\tilde{e}_L}\geq m_{\widetilde 
{B}}$. The pair production cross sections for the slepton are obtained 
using the \texttt{Resummino} computer code at the NLL 
level for $\sqrt{s}=$ $\SI{13}{\tera\electronvolt}$ center-of-mass energy at the
LHC. 

\begin{figure}[htbp!]
    \centering
    \includegraphics[width=0.6\textwidth]{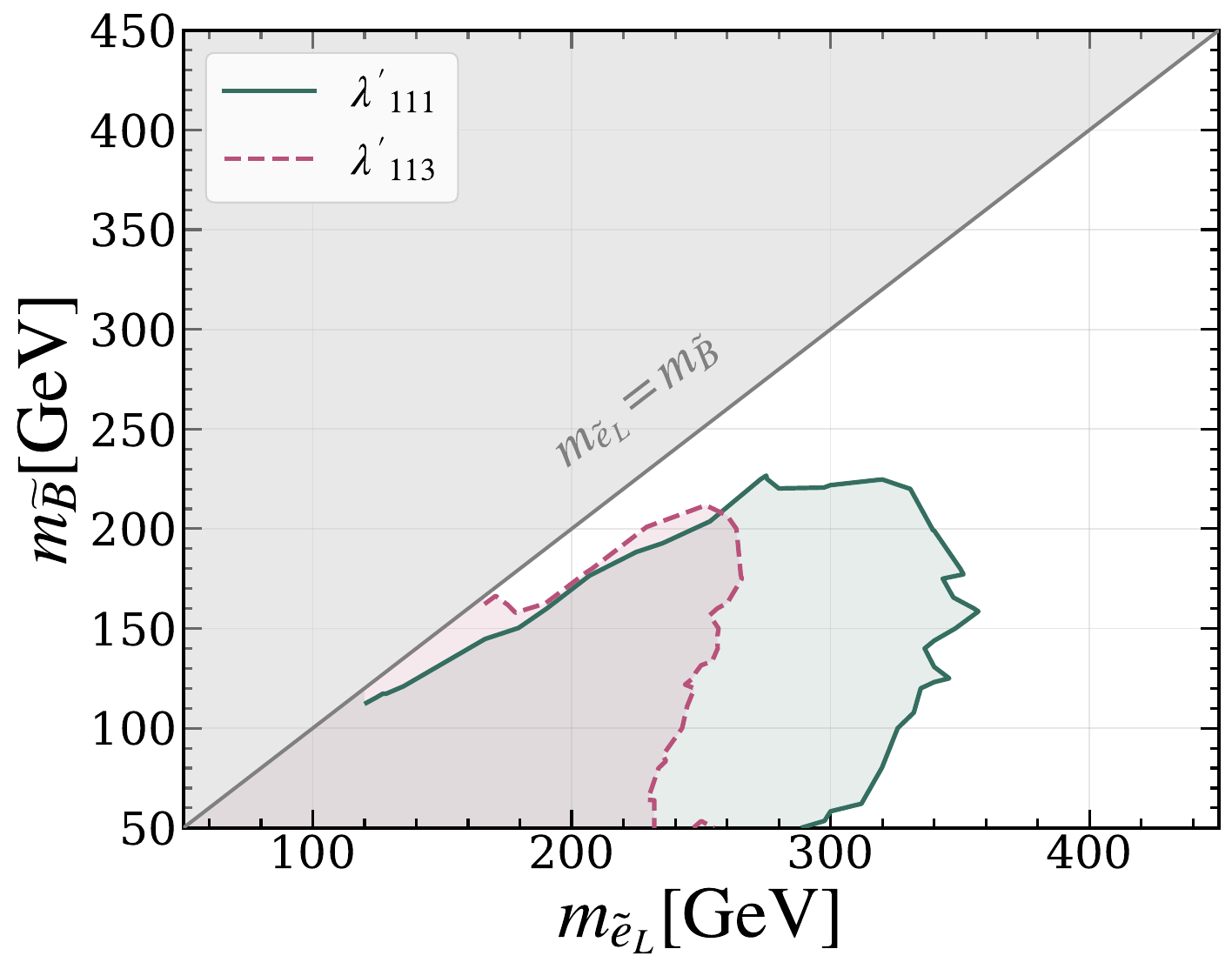}
    \caption{Results for the cascade decay of an NLSP selectron, $\tilde{e}_L$,  to an LSP bino, $\widetilde{B}$. The gray region depicts the kinematically inaccessible region. A list of sensitive searches can be found in Table~\ref{tab:sleptontobino} of App.~\ref{sec:appendix-A2}.}
    \label{fig:sleptontobino}
\end{figure}

The results for this cascade are shown in Fig.~\ref{fig:sleptontobino}. 
The scan was performed for the couplings $\lam'_{111},\,\lam'_{113},\, 
\lam'_{131}$ and $\lam'_{133}$, respectively. However, we only show
results for the two couplings $\lam'_{111}\,$ and $\lam'_{113}$. For the couplings 
$\lam'_{131}$ and $\lam'_{133}$, a clear exclusion contour could not be 
drawn from the \texttt{CheckMATE\,2} results, due to the low 
expected statistics. We provide a scatter plot of the inconclusive contour obtained for these two couplings in 
App.~\ref{sec:appendix-A3}, Fig.~\ref{fig:slepton_bino_scatter}. We also list the searches that contribute sensitivity to this benchmark across 
all couplings, ordered by relevance, in Table~\ref{tab:sleptontobino} 
in App.~\ref{sec:appendix-A2}.

For completeness, we also performed a scan for the 
cascade decay of an NLSP stau, $\tilde{\tau}_L$, to an LSP 
bino, $\widetilde{B}$, for the couplings $\lam'_{311},\,\lam'_{313},\, 
\lam'_{331}$ and $\lam'_{333}$. However, we did not obtain an exclusion contour for this benchmark from any currently 
implemented searches in \texttt{CheckMATE\,2}.

\subsubsection*{Wino NLSPs}
For scenarios with a wino-like NLSP, $\widetilde{W}$, cascade decaying to a bino-like LSP, $\widetilde{B}$, we impose a minimum wino-bino mass splitting of $\SI{250}{\giga\electronvolt}$. This ensures that the LSP and NLSP remain dominantly bino- and wino-like, respectively~\cite{Profumo:2017ntc}, while keeping the decays involving $W^\pm$, $Z$, and $h$ bosons kinematically accessible. The mass splitting also determines the kinematics of these bosons in the final state.
The pair production cross sections for the 
$\widetilde{W}$ are obtained using the 
\texttt{Resummino} computer code at the NLO+NLL order at $\sqrt{s}=$ $\SI{13}{\tera\electronvolt}$, for the production channels $pp\to \tilde{\chi}^0_2\tilde 
{\chi}^{\pm}_1$ and $pp \to \tilde{\chi}^{+}_1\tilde{\chi}^{-}_1$. We assume equal branching fractions for the $\tilde{\chi}^0_2$ NLSP decay into the $Z+\widetilde{B}$ and $h+\widetilde{B}$ channels.

\begin{figure}[htbp!]
    \centering
    \includegraphics[width=0.6\textwidth]{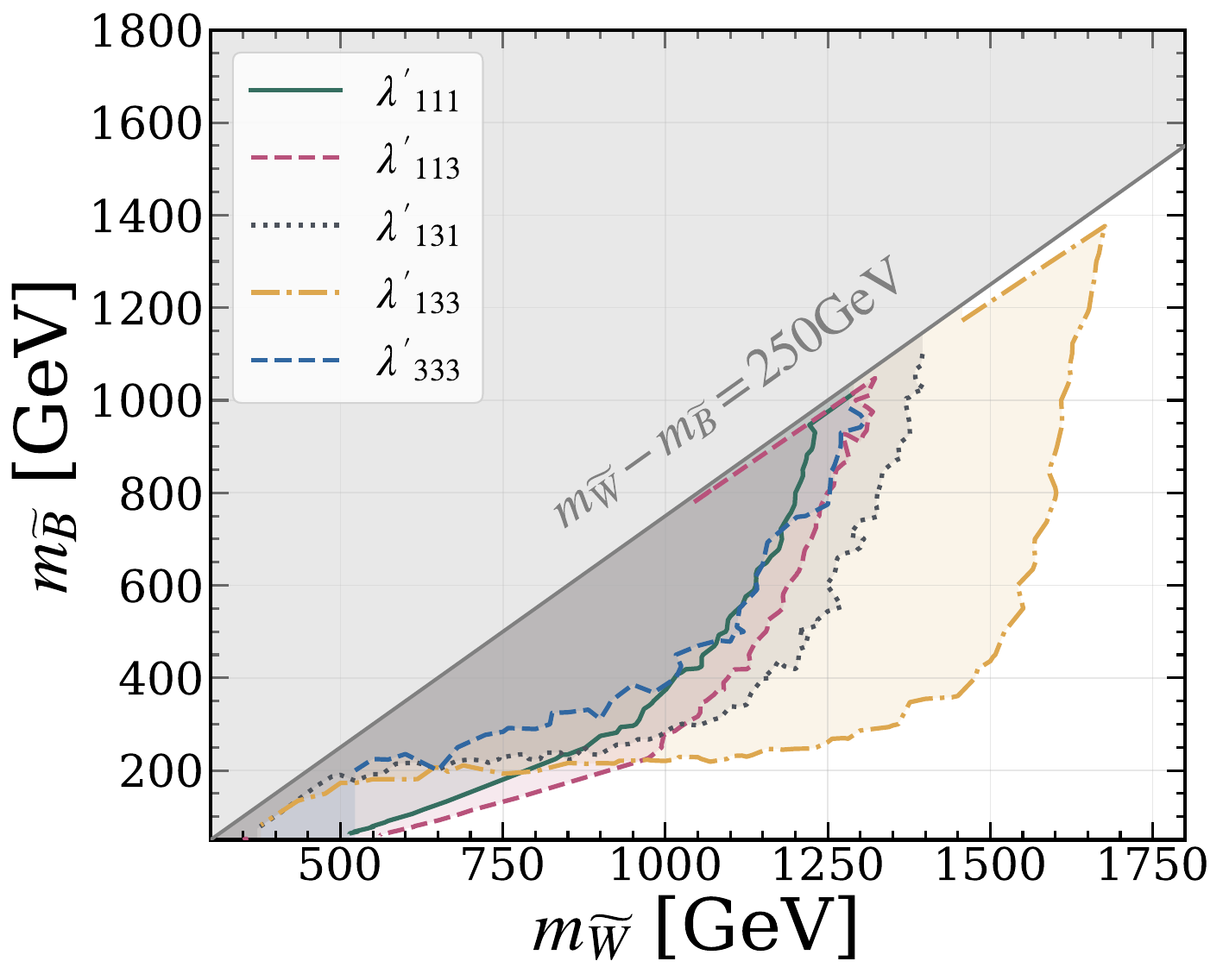}
    \caption{Results for the cascade decay of an NLSP wino, 
    $\widetilde{W}$, to an LSP bino, $\widetilde{B}$. We obtain no 
    exclusion for couplings not shown in the plot. The gray region 
    corresponds to the region with $m_{\widetilde{W}}-m_{\widetilde{B}} 
    \leq\SI{250}{\giga\electronvolt}$, which is not considered in these scenarios. A list of sensitive searches can be found in 
    Table~\ref{tab:winotobino} of App.~\ref{sec:appendix-A2}.}
    \label{fig:winotobino}
\end{figure}

The results for this cascade are presented in 
Fig.~\ref{fig:winotobino}, in the ($m_{\widetilde 
{W}}, m_{\widetilde{B}}$) mass plane. The couplings not shown in the plot 
do not yield any exclusion in \texttt{CheckMATE\,2}, namely $\lam'_{311},\, \lam'_{313}$ and $\lam'_{331}$. As in previous benchmarks, $\lam'_{133}$ achieves the largest exclusion in the mass plane, reaching up to $\SI{1650}{\giga\electronvolt}$ in $m_{\widetilde{W}}$ for a bino-like LSP of mass of about $\SI{1300}{\giga\electronvolt}$. We again list the searches that 
contribute sensitivity to this benchmark across all couplings, ordered by 
relevance, in Table~\ref{tab:winotobino}, in App.~\ref{sec:appendix-A2}.

\section{Conclusions}
\label{sec:conclusion}

Supersymmetry with broken R-parity offers a wide range of potential 
signatures at the LHC. As such it is an ideal testing ground for the 
coverage of beyond the Standard Model searches at the LHC. This work extends the framework introduced in
Ref.~\cite{Dreiner_2023}, which initially systematically
investigated the sensitivity of existing ATLAS and CMS searches at 
the LHC to a specific subset of supersymmetric R-parity violating 
operators, namely the nine $LL\Bar{E}$ operators. This was based on a classification of a wide range of signatures from the full set of 
operators in terms of a small representative set of operators, covering all possible signatures. All possible LSPs were allowed. For the 
production, only direct LSP pair production or the production of an NLSP particle with at most a one-step decay to the LSP were allowed. The R-parity violating couplings were assumed small, such that only the LSP could 
decay via an R-parity violating operator. The resulting coverage at the 
LHC was determined. For the $LL\bar E$ couplings, Ref.~\cite{Dreiner_2023} obtained strong exclusion
limits for all possible LSPs, with the colored sector showing the best coverage.

Subsequently the corresponding 
investigation of the signatures resulting from the dominant nine 
$\bar{U}\bar{D}\bar{D}$ operators was performed in Ref.~\cite{Dreiner:2025kfd}. We find that gluino and squark LSPs are generally well covered, whereas significant gaps remain for slepton, wino, and higgsino LSPs decaying via $\bar U\bar D\bar D$ couplings. Nevertheless, some squark LSP scenarios remain only weakly constrained; for example, the limit on a $\tilde{d}_R$ LSP decaying via $\lambda''_{312}$ lies below 500\,GeV. 

In this 
paper we have performed the corresponding analysis of the 27 R-parity violating $LQ\bar{D}$ operators in the RPV-MSSM. We performed a complete treatment of the phenomenology
with minimal assumptions, including again the assumption
of small $LQ\bar{D}$ R-parity violating couplings. We divided the $LQ\bar{D}$ couplings into eight classes based on the resulting final states at the LHC. We then 
identified one coupling in each class as the representative benchmark coupling, for which we then analysed
the phenomenology. For this we consider in turn each possible 
LSP. The LSPs themselves can be pair produced at the LHC, or be produced
through a cascade decay of the NLSP via gauge couplings. Due to the
small R-parity violating coupling, only the LSP can decay via the 
$LQ\bar D$ operators. We then identified the final states for 
each respective benchmark coupling when the LSP decays through an 
$LQ\bar{D}$ coupling. We then performed a recasting of existing 
ATLAS and CMS searches for the final states arising from LSPs decaying 
through $LQ\bar{D}$ couplings, using the recasting tool 
\texttt{CheckMATE\,2}. 

We have also identified potential ATLAS and CMS searches 
which could be relevant for the broad classes of final states arising 
from $LQ\bar{D}$ operators. We have presented an updated 
list of searches which could be sensitive to $LQ\bar {D}$ couplings in 
Sect.~\ref{sec:relevant_analyses}. We then tested a selected set 
of our list of LSP final states with the current version of 
\texttt{CheckMATE\,2}, with ATLAS and CMS searches at $\sqrt{s}=\SI{13} 
{\tera\electronvolt}$. Benchmark scenarios are chosen such that the LSP 
decays through a zero-step or one-step decay. We have 
presented our results in
Figs.~\ref{fig:gluino_direct}-\ref{fig:ewino_direct} for direct LSP 
production, and Figs.~\ref{fig:gluinotobino}-\ref{fig:winotobino} for cascade LSP production. 

We summarise as follows:
\begin{itemize}
    \item We found that a majority of the searches listed  in Sect.~\ref{sec:relevant_analyses} are not implemented in \texttt{CheckMATE\,2}. We expect that the implementation of these searches should provide a better coverage of $LQ\bar{D}$ final states.
    
    \item The recent implementation of full and simplified 
    likelihoods in \texttt{CheckMATE\,2} improves the coverage of 
    $LQ\bar{D}$ final states, compared to when one uses just the signal 
    region based searches implemented in \texttt{CheckMATE\,2}. We have observed that the coupling $\lam'_{133}$ yields the best mass 
    exclusions for most LSPs and for direct and indirect 
    production, owing to the good 
    coverage of signatures with $t-$ and/or $b-$ jets in the
    final state.
    
    \item The colored sector is well covered for $LQ\bar{D}$ couplings. For the direct production of the gluino LSP, we achieve mass exclusions $> \SI{2}{\tera\electronvolt}$ in all scenarios. For most direct production benchmarks of squark LSPs, mass exclusions are above $\SI{1}{\tera\electronvolt}$, lying between $\SI{0.8}{\tera\electronvolt}-\SI{1.8}{\tera\electronvolt}$.
    
    \item We have obtained no exclusion for direct 
    production of slepton LSPs. This can mainly be attributed to their small expected pair production cross sections at the LHC. However, new searches, especially with $4j$ final states should greatly improve coverage in this case. The searches for multijet final states targeting the coloured sector often impose high thresholds on jet energies. For sleptons having such multijet final states, searches with a lower threshold on jets are expected to provide sensitivity. 
    
    \item We have showed that the electroweakinos are also well
    covered for $LQ\bar{D}$ couplings for certain benchmarks, except for
    the bino. Wino and higgsino LSPs can be excluded up to a TeV for
    some direct LSP production scenarios. Benchmarks with $\tau/\nu_
    {\tau}$ in the final state still only have mass exclusions up to a 
    few hundred GeV. No exclusion was found for the benchmarks of wino
    LSPs with $\lam'_{331}$ and higgsino LSPs with $\lam'_{313},\lam'
    _{331}$. In the future, experimentalists can target final
    states with the decay LSP $\to \tau/\nu_{\tau} +X$ to improve
    coverage. 
    
    \item Due to the small production cross sections of the bino LSP, we did not perform a numerical analysis for direct bino
    LSP pair production. However, we considered cascade
    decay scenarios where the bino is produced through the decay
    of an NLSP. In all our presented benchmarks, we have
    shown exclusions in the NLSP-LSP mass plane. As NLSPs, we considered: $\tilde{g},\,\tilde{q}_L,\, 
    \tilde{t}_1,\,\tilde{e}_L$, and $\widetilde{W}$. The largest 
    exclusion is achieved in the $\tilde{g}\to
    \widetilde{B}+X$ scenario. The $\tilde{e}_L\to\widetilde{B}+X$ 
    scenario has the smallest coverage. Here an exclusion contour reaching up to a maximum of
    about $\SI{320}{\giga\electronvolt}$ in the $\tilde{e}_L$ mass
    for a $\widetilde{B}$ mass of $\SI{200}{\giga\electronvolt}$ 
    in the $\tilde{e}_L-\widetilde{B}$ mass plane was found for the 
    $\lam'_{111}$ coupling. No conclusive exclusion contours could be 
    drawn for the $\lam'_{131}$ and $\lam'_{133}$ couplings. We have observed a gap for stau NLSP $\rightarrow$ bino LSP, 
    where we obtain no exclusion. Once again, this can be attributed to 
    $\tau$'s in the final state, for which efficiencies are much 
    lower. 
    
    
    \item Given the limited coverage of $\tau$-rich final states, we highlight an important gap in the current \texttt{CheckMATE\,2} analysis library. Several searches discussed in Section~\ref{sec:relevant_analyses} employ dedicated hadronic-$\tau$ identification and define signal regions explicitly optimised for final states containing one or more $\tau$-leptons, but are not presently implemented in \texttt{CheckMATE\,2}. In particular, these include the LHC searches \texttt{cms\_1910\_12932}~\cite{CMS:2019lrh}, \texttt{cms\_2012\_04178}~\cite{CMS:2020wzx}, \texttt{atlas\_2103\_11684}~\cite{ATLAS:2021yyr}, \texttt{cms\_2106\_14246}~\cite{CMS:2021cox}, \texttt{cms\_2208\_09700}~\cite{CMS:2022cpe}, and \texttt{atlas\_2507\_00296}~\cite{ATLAS:2025wln}.
    Implementing these analyses could substantially improve the sensitivity of reinterpretation studies to models characterised by $\tau$-dominated decay chains.
\end{itemize}

Our findings point towards requiring a further joint effort from 
theorists and experimentalists in order to probe and improve the 
coverage of the $LQ\bar{D}$ operators of RPV-MSSM at the LHC. 
New experimental searches, along with an implementation of existing 
searches into recasting frameworks are required.

This is the third in a series of papers. Given the framework outlined
in Ref.~\cite{Dreiner_2023}, we have classified a general set of signatures
for R-parity violating models within the MSSM. This is not as general as 
the full set of signatures given in \cite{Dreiner:2012wm}, which involve 
all possible spectra, as well as potential R-parity violating decays within
the cascade. Instead in these scenarios we have limited ourselves to all 
possible LSPs and related the NLSPs. We have considered the pair production
only of the LSPs or the NLSPs. In Ref.~\cite{Dreiner_2023} we have 
discussed the signatures for the case of a single dominant $LL\bar E$ 
operator, of which there are nine. In Ref.~\cite{Dreiner:2025kfd} we have 
analogously discussed the case of the $\bar U\bar D\bar D$ operators, of 
which there are also nine. Here in this paper we have then discussed the 
more extensive case of the 27 $LQ\bar D$ operators. In all cases we have 
considered only prompt decays. 
Overall, while the colored sector is generally well covered across all three RPV operators, $LL\bar E$, $\bar U\bar D\bar D$, and $LQ\bar D$, the experimental coverage of slepton and electroweakino sectors progressively weakens from $LL\bar E$ to $LQ\bar D$, and is particularly limited for $\bar U\bar D\bar D$. What remains to be considered is the case
of a long-lived LSP, with an observable decay length 
\cite{Domingo:2023dew,Bhattacherjee:2023kxw,Cottin:2025avd,Dreiner:2026ppi}. One could also consider the case of bi-linear 
R-parity violation, which leads to alternative signatures 
\cite{deCampos:2007bn,Choudhury:2026imz}.

\medskip

\section*{Acknowledgements}

The authors would like to thank Krzysztof Rolbiecki for useful discussions on the likelihood implementation of searches in \texttt{CheckMATE\,2}.
NS is supported by the U.S. Department of Energy, Office of Science, Office of High Energy Physics under Award Number DE-SC0011845. 

\appendix
\section{Appendix}
\label{sec:appendix}

\subsection{Supplementary Tables for Direct Production}
\label{sec:appendix-A1}
\begin{table}[H]
    \centering
    \begin{tabular}{|c|c|c|}
    \hline\hline
        Coupling & Exclusion& Sensitive Search and Signal Region\\
        \hline\hline
         $\lam'_{111}$&2260 GeV &\texttt{atlas\_2101\_01629, combined}\\
         $\lam'_{113}$&2280 GeV &\texttt{atlas\_2101\_01629, combined}\\
         $\lambda'_{131}$&2300 GeV &\texttt{atlas\_2101\_01629, combined}\\
         $\lambda'_{133}$& 2640 GeV &\texttt{atlas\_2211\_08028, CC\_Gtt\_1L\_M1}\\
         $\lambda'_{311}$& 2240 GeV &\texttt{atlas\_2010\_14293, MB-GGd}\\
         $\lambda'_{313}$& 2220 GeV &\texttt{atlas\_2010\_14293, MB-GGd}\\
         $\lambda'_{331}$& 2160 GeV &\texttt{atlas\_2010\_14293, MB-GGd}\\
         $\lambda'_{333}$& 2300 GeV &\texttt{atlas\_2211\_08028, CC\_Gtt\_1L\_M1}\\
         \hline\hline
    \end{tabular}
    \caption{Results for direct production of the gluino $\tilde{g}$ LSP. We list the ATLAS/CMS search and the corresponding signal region that provide the best exclusion limit in each case obtained using \texttt{CheckMATE\,2}, as shown in Fig.~\ref{fig:gluino_direct}.
    }
    \label{tab:direct_gluino}
\end{table}

\begin{table}[H]
    \centering
    \begin{tabular}{|c|c|c|}
    \hline\hline
        Coupling & Exclusion& Sensitive Search and Signal Region\\
        \hline\hline
         $\lambda'_{111}$&1260 GeV &\texttt{atlas\_2010\_01629, MB-SSd}\\
         $\lambda'_{111}$&1800 GeV &\texttt{CMS-PAS-EXO-24-019} (scaling)\\
         $\lambda'_{113}$&1240 GeV &\texttt{atlas\_2211\_08028, CC\_Gtt\_1L\_M1}\\
         $\lambda'_{113}$&1800 GeV &\texttt{CMS-PAS-EXO-24-019} (scaling)\\
         $\lambda'_{311}$& 1260 GeV &\texttt{atlas\_2010\_14293, low}\\
         $\lambda'_{313}$& 1220 GeV &\texttt{cms\_sus\_19\_005, low}\\
         \hline\hline
    \end{tabular}
    \caption{Results for direct production of the $\tilde{q}_L$ LSP. We list the ATLAS/CMS search and the corresponding signal region that provide the best exclusion limit in each case obtained using \texttt{CheckMATE\,2}, as shown in Fig.~\ref{fig:squark_direct}. We also show exclusion obtained from the cross-section scaling of the results from \texttt{CMS-PAS-EXO-24-019}~\cite{CMS-PAS-EXO-24-019}.}
    \label{tab:qL_direct}
\end{table}

\begin{table}[H]
    \centering
    \begin{tabular}{|c|c|c|}
    \hline\hline
        Coupling & Exclusion& Sensitive Search and Signal Region\\
        \hline\hline
         $\lambda'_{131}$&660 GeV &\texttt{atlas\_2209\_03935, SR-1J}\\
         $\lambda'_{131}$&1800 GeV &\texttt{CMS-PAS-EXO-24-019} (scaling)\\
         $\lambda'_{133}$&1240 GeV &\texttt{atlas\_2211\_08028, CC\_Gtt\_1L\_M1}\\
         $\lambda'_{133}$&1800 GeV &\texttt{CMS-PAS-EXO-24-019} (scaling)\\
         $\lambda'_{331}$& 880 GeV &\texttt{atlas\_1911\_06660, combined}\\
         $\lambda'_{333}$& 1000 GeV &\texttt{atlas\_2211\_08028, CC\_Gtt\_1L\_M1}\\
         \hline\hline
    \end{tabular}
    \caption{Results for direct production of the stop $\tilde{t}_L$ LSP. We list the ATLAS/CMS search and the corresponding signal region that provide the best exclusion limit in each case obtained using \texttt{CheckMATE\,2}, as shown in Fig.~\ref{fig:squark_direct}. We also show exclusion obtained from the cross-section scaling of the results from \texttt{CMS-PAS-EXO-24-019}~\cite{CMS-PAS-EXO-24-019}.}
    \label{tab:tL_direct}
\end{table}

\begin{table}[H]
    \centering
    \begin{tabular}{|c|c|c|}
    \hline\hline
        Coupling & Exclusion& Sensitive Search and Signal Region\\
        \hline\hline
         $\lambda'_{113}$&1280 GeV &\texttt{atlas\_2101\_01629, combined}\\
         $\lambda'_{113}$&1600 GeV &\texttt{CMS-PAS-EXO-24-019} (scaling)\\
         $\lambda'_{133}$&1460 GeV &\texttt{atlas\_2211\_08028, CC\_Gtt\_1L\_M1}\\
         $\lambda'_{133}$&1600 GeV &\texttt{CMS-PAS-EXO-24-019} (scaling)\\
         $\lambda'_{313}$& 1100 GeV &\texttt{atlas\_2010\_14293, MB-SSd}\\
         $\lambda'_{333}$& 1140 GeV &\texttt{atlas\_2004\_14060, SRA-T0}\\
         \hline\hline
    \end{tabular}
    \caption{Results for direct production of the sbottom  $\tilde{b}_R$ LSP. We list the ATLAS/CMS search and the corresponding signal region that provide the best exclusion limit in each case obtained using \texttt{CheckMATE\,2}, as shown in Fig.~\ref{fig:squark_direct}. We also show exclusion obtained from the cross-section scaling of the results from \texttt{CMS-PAS-EXO-24-019}. The results for $\lambda'_{ij3}$ coupling for the $\tilde{b}_R$ LSP also apply to the $\tilde{d}_R$ LSP case for $\lambda'_{ij1}$ coupling.}
    \label{tab:bR_direct}
\end{table}

\begin{table}[H]
    \centering
    \begin{tabular}{|c|c|c|}
    \hline\hline
        Coupling & Exclusion& Sensitive Search and Signal region\\
        \hline\hline
         $\lambda'_{131}$&1280 GeV &\texttt{atlas\_2010\_14293, MB-SSd}\\
         $\lambda'_{133}$&1240 GeV &\texttt{cms\_sus\_19\_005, low}\\
         $\lambda'_{331}$& 1220 GeV &\texttt{cms\_sus\_19\_005, low}\\
         $\lambda'_{333}$& 1240 GeV &\texttt{cms\_sus\_19\_005, low}\\
         \hline\hline
    \end{tabular}
    \caption{Results for the direct production of sbottom $\tilde{b}_L$ LSP. We list the ATLAS/CMS search and corresponding signal region for the exclusion limits obtained using \texttt{CheckMATE\,2} shown in Fig.~\ref{fig:squark_direct}.}
    \label{tab:bL_direct}
\end{table}

\begin{table}[ht!]
    \centering
    \begin{tabular}{|c|c|c|}
    \hline\hline
        Coupling & Exclusion& Sensitive Search and Signal region\\
        \hline\hline
         $\lambda'_{111}$&1000 GeV &\texttt{atlas\_2101\_01629, 6J\_disc\_squark}\\
         $\lambda'_{113}$&1175 GeV &\texttt{atlas\_2101\_01629, combined}\\
         $\lambda'_{131}$&1200 GeV &\texttt{atlas\_2101\_01629, combined}\\
         $\lambda'_{133}$& 1575 GeV &\texttt{atlas\_2211\_08028, CC\_Gtt\_1L\_M1}\\
         $\lambda'_{311}$& 400 GeV &\texttt{atlas\_2004\_10894, Cat12}\\
         $\lambda'_{313}$& 750 GeV &\texttt{atlas\_2004\_14060, SRA-T0}\\
         $\lambda'_{331}$& -- & no exclusion\\
         $\lambda'_{333}$& 800 GeV &\texttt{atlas\_2211\_08028, CC\_Gtt\_1L\_M2}\\
         \hline\hline
    \end{tabular}
    \caption{Results for the direct production of wino $\widetilde{W}$ LSPs. We list the ATLAS/CMS search and corresponding signal region for the exclusion limits obtained using \texttt{CheckMATE\,2} shown in Fig.~\ref{fig:ewino_direct}. For the coupling $\lambda'_{331}$ we obtain no mass exclusion limit.}
    \label{tab:W_direct}
\end{table}

\begin{table}[ht!]
    \centering
    \begin{tabular}{|c|c|c|}
    \hline\hline
        Coupling & Exclusion& Sensitive Search and Signal region\\
        \hline\hline
         $\lambda'_{111}$&900 GeV &\texttt{atlas\_2101\_01629,combined}\\
         $\lambda'_{113}$&950 GeV &\texttt{atlas\_2101\_01629, combined}\\
         $\lambda'_{131}$&950 GeV &\texttt{atlas\_2101\_01629, combined}\\
         $\lambda'_{133}$& 1350 GeV &\texttt{atlas\_2211\_08028, CC\_Gtt\_1L\_M1}\\
         $\lambda'_{311}$& 300 GeV &\texttt{cms\_sus\_16\_039, SR\_K03}\\
         $\lambda'_{313}$& -- & no exclusion\\
         $\lambda'_{331}$& -- & no exclusion\\
         $\lambda'_{333}$& 150 GeV &\texttt{atlas\_1807\_07447, 3b1j}\\
         \hline\hline
    \end{tabular}
    \caption{Results for the direct production of higgsino $\widetilde{H}$ LSPs. We list the ATLAS/CMS search and corresponding signal region for the exclusion limits obtained using \texttt{CheckMATE\,2} shown in Fig.~\ref{fig:ewino_direct}. For the couplings $\lambda'_{313}$ and $\lambda'_{331}$ we obtain no mass exclusion limit.}
    \label{tab:H_direct}
\end{table}

\newpage
\subsection{Supplementary Tables for Cascade Production}
\label{sec:appendix-A2}

In this section, we list all sensitive searches for the cascade decay benchmark results presented in section~\ref{sec:cascadedecay}. The searches are ordered by relevance, where they appear in decreasing order of the number of mass points they exclude. This is inclusive of both approaches of \texttt{CheckMATE\,2}, based on $r-$ and CL$_s$ values. Since the number of corresponding signal regions across all couplings is too large, we omit that information here.

\begin{table}[H]
    \centering
    \begin{tabular}{|c|c|}
    \hline\hline
    Cascade  &   Sensitive Searches \\
    \hline\hline
    \multirow{5}{*}{$\widetilde{g}\rightarrow \widetilde{B}$} & \texttt{atlas\_2101\_01629, cms\_sus\_19\_005}\\
    &\texttt{atlas\_2211\_08028, cms\_1908\_04722}\\
    &\texttt{atlas\_2010\_14293, atlas\_2106\_09609}\\
    &\texttt{atlas\_1712\_02332, atlas\_1807\_07447}\\
    &\texttt{atlas\_2004\_10894}\\

    \hline\hline  
    
    \end{tabular}
    \caption{Sensitive searches for the results of cascade decay $\tilde{g}\rightarrow\widetilde{B}$ for all couplings (shown in Fig.~\ref{fig:gluinotobino}), ordered by relevance.}
    \label{tab:gluinotobino}
\end{table}

\begin{table}[H]
    \centering
    \begin{tabular}{|c|c|}
    \hline\hline
    Cascade  &   Sensitive Searches \\
    \hline\hline
    \multirow{5}{*}{$\widetilde{q}_L\rightarrow \widetilde{B}$} & \texttt{atlas\_2101\_01629, cms\_1908\_04722, atlas\_2106\_09609}\\
&\texttt{atlas\_2010\_14293, atlas\_1807\_07447, cms\_sus\_19\_005}\\
&\texttt{atlas\_2004\_10894, atlas\_2211\_08028, atlas\_1706\_03731}\\
&\texttt{atlas\_1712\_02332, atlas\_1709\_04183, atlas\_2004\_14060}\\
&\texttt{atlas\_conf\_2016\_050, cms\_2107\_13201, cms\_sus\_16\_039}\\
\hline
\multirow{2}{*}{$\widetilde{t}_1\rightarrow \widetilde{B}$} & \texttt{atlas\_2211\_08028, atlas\_2101\_01629}\\
&\texttt{atlas\_2006\_05880, 
atlas\_1908\_03122}\\

    \hline\hline  
    
    \end{tabular}
    \caption{Sensitive searches for the results of cascade decay $\tilde{q}_L/\tilde{t}_1$ $\rightarrow\widetilde{B}$ for all couplings (shown in fig.~\ref{fig:squarktobino}), ordered by relevance. }
\label{tab:squarktobino}
\end{table}

\begin{table}[H]
    \centering
    \begin{tabular}{|c|c|}
    \hline\hline
    Cascade  &   Sensitive Searches \\
    \hline\hline
    \multirow{3}{*}{$\widetilde{e}_L\rightarrow \widetilde{B}$} & \texttt{atlas\_conf\_2016\_096, atlas\_2209\_13935}\\
&\texttt{atlas\_1706\_03731, atlas\_1908\_08215} \\
&\texttt{atlas\_2106\_01676, 
cms\_sus\_16\_039}\\
                                     
    \hline\hline  
    
    \end{tabular}
    \caption{Sensitive searches for the results of cascade decay $\tilde{e}_L\rightarrow\widetilde{B}$ for all couplings (shown in Fig.~\ref{fig:sleptontobino}), ordered by relevance.}
\label{tab:sleptontobino}
\end{table}

\begin{table}[H]
    \centering
    \begin{tabular}{|c|c|}
    \hline\hline
    Cascade  &   Sensitive Searches \\
    \hline\hline
    \multirow{3}{*}{$\widetilde{W}\rightarrow \widetilde{B}$} & \texttt{atlas\_2211\_08028, atlas\_2101\_01629}\\
&\texttt{atlas\_2106\_09609, atlas\_2006\_05880}\\
&\texttt{atlas\_1706\_03731}\\

    \hline\hline  
    
    \end{tabular}
    \caption{Sensitive searches for the results of cascade decay $\widetilde{W}\rightarrow\widetilde{B}$ for all couplings (shown in Fig.~\ref{fig:winotobino}), ordered by relevance. }
\label{tab:winotobino}
\end{table}

\subsection{Supplementary Figures}
\label{sec:appendix-A3}
\begin{figure}[H]
    \centering
    \begin{minipage}{0.49\textwidth}
    \includegraphics[width=\textwidth]{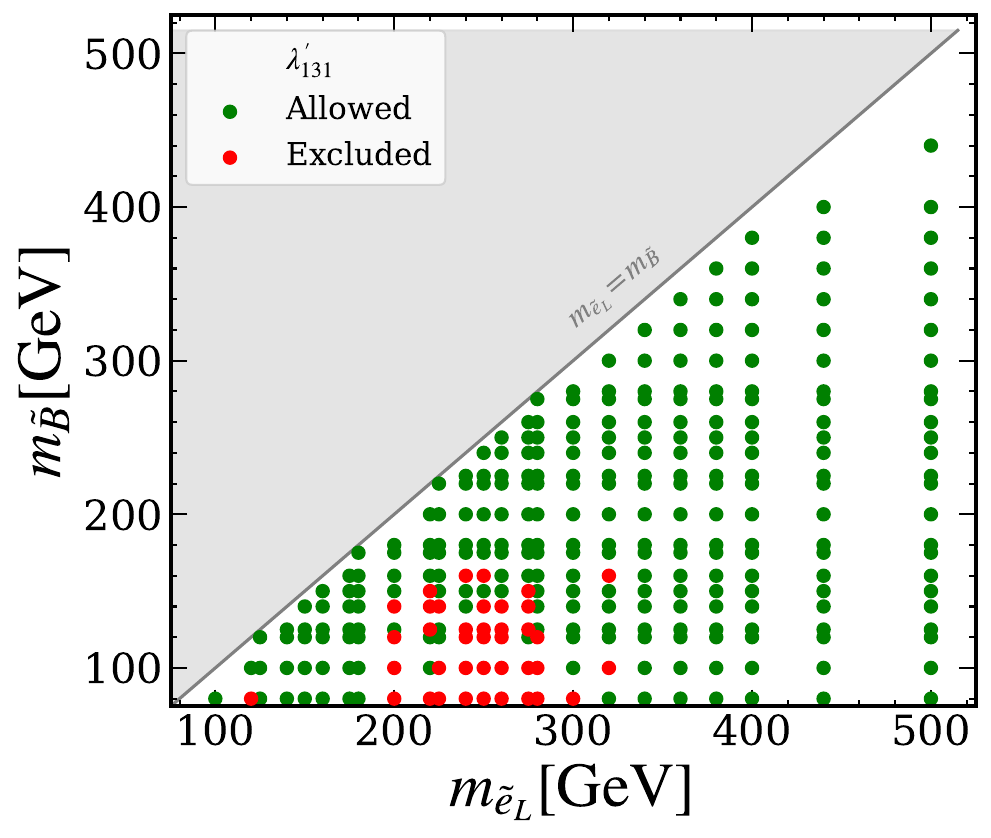}
    \end{minipage}
    \begin{minipage}{0.49\textwidth}
    \centering
    \includegraphics[width=\textwidth]{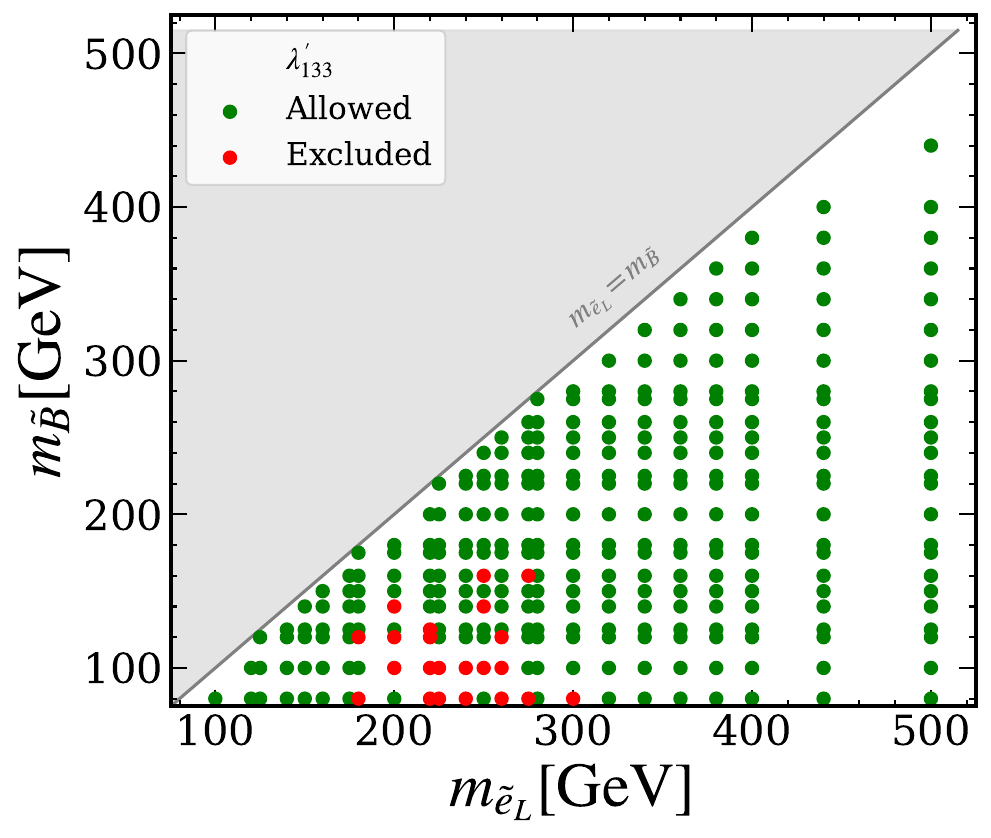}
    \end{minipage}
    \caption{Scatter plots for the results of cascade decay $\tilde{e}_L \rightarrow \widetilde{B}$. The red points are excluded by \texttt{CheckMATE\,2}.}
    \label{fig:slepton_bino_scatter}
\end{figure}

As mentioned in Sec.~\ref{sec:sleptontobino}, a conclusive well-defined contour could not be drawn for the results of the cascade decay $\tilde{e}_L \rightarrow \widetilde{B}$ for the couplings $\lambda'_{131}$ and $\lambda'_{133}$. We show the scan results in Fig.~\ref{fig:slepton_bino_scatter}, where each point was scanned with 50,000 generated events. The red dots correspond to points excluded by the implemented searches in \texttt{CheckMATE\,2}, and the green points are allowed by \texttt{CheckMATE\,2}. No definite connected area emerges from the points that are excluded. We attribute this to the low production cross section of sleptons.
Some of the isolated excluded points arise from changes in the most sensitive analysis. For example, the mass point ($m_{\tilde{e}_L}=$ 120\,GeV, $m_{\widetilde{B}}=$ 80\,GeV) for $\lam'_{131}$ is excluded by the search \texttt{atlas\_2209\_13935}, while the points around it are allowed with the most sensitive search being \texttt{cms\_sus\_16\_039}. 


\bibliography{ref}
\end{document}